\documentclass[11pt]{article}
\usepackage{algpseudocode}
\usepackage[final]{acl}

\usepackage{times}
\usepackage{latexsym}
\usepackage{listings}
\usepackage{fvextra}
\usepackage{placeins}
\usepackage{enumitem}
\usepackage[most]{tcolorbox}
\usepackage{xcolor}
\usepackage{multirow}
\usepackage{amsmath,amssymb,amsfonts}
\usepackage{algorithm}
\usepackage{subcaption}
\usepackage{wrapfig} 
\usepackage{booktabs}
\usepackage{adjustbox}

\usepackage{tcolorbox}
\definecolor{lightgray}{gray}{0.9}
\definecolor{headergray}{gray}{0.8}
\definecolor{darkgray}{gray}{0.6}
\usepackage{colortbl}
\usepackage{pifont}
\usepackage[T1]{fontenc}

\usepackage[utf8]{inputenc}

\usepackage{microtype}

\usepackage{inconsolata}

\usepackage{graphicx}

\title{CollabCoder: Plan-Code Co-Evolution via Collaborative Decision-Making for Efficient Code Generation}

\author{
  Duy Tung Doan$^{1,2}$\thanks{Equal contribution.}
  \quad
  Quang Huy Phung$^{1}$\footnotemark[1]
  \quad
  Ngoc Dung Nguyen$^{1}$
  \quad
  Khac-Hoai Nam Bui$^{1}$\thanks{Corresponding author.} \\
  $^{1}$Viettel AI, Viettel Group, Vietnam \\
  $^{2}$Hanoi University of Science and Technology, Vietnam \\
  \texttt{\{tungdd11, huypq51, dungnn7, nambkh\}@viettel.com.vn} 
}

\begin{document}
\maketitle
\begin{abstract}
Automated code generation remains a persistent challenge in software engineering, as conventional multi-agent frameworks are often constrained by static planning, isolated execution, high computational overhead, and limited adaptability to complex tasks. This paper introduces CollabCoder\footnote{The source code is publicly available at \url{https://github.com/ihbkaiser/CollabCoder}.}, a novel Plan-Code Co-Evolution framework that improves code generation through dynamic multi-agent collaboration. The core idea is to design a collaborative decision-making module between the plan agent and the code agent to decide which should be executed for the debugging process. Extensive experiments on widely used benchmarks demonstrate that CollabCoder consistently improves code quality and robustness across tasks. Importantly, CollabCoder achieves performance comparable to or exceeding current state-of-the-art methods while reducing computational overhead, with efficiency gains becoming more pronounced as benchmark difficulty increases. 
On the more challenging LiveCodeBench and xCodeEval benchmarks, our approach improves performance by 11-20\% over strong baselines while reducing the number of API calls by an average of 4-10 per execution.

\end{abstract}

\section{Introduction}
\begin{figure}[h]
    \centering
    \includegraphics[width=\linewidth]{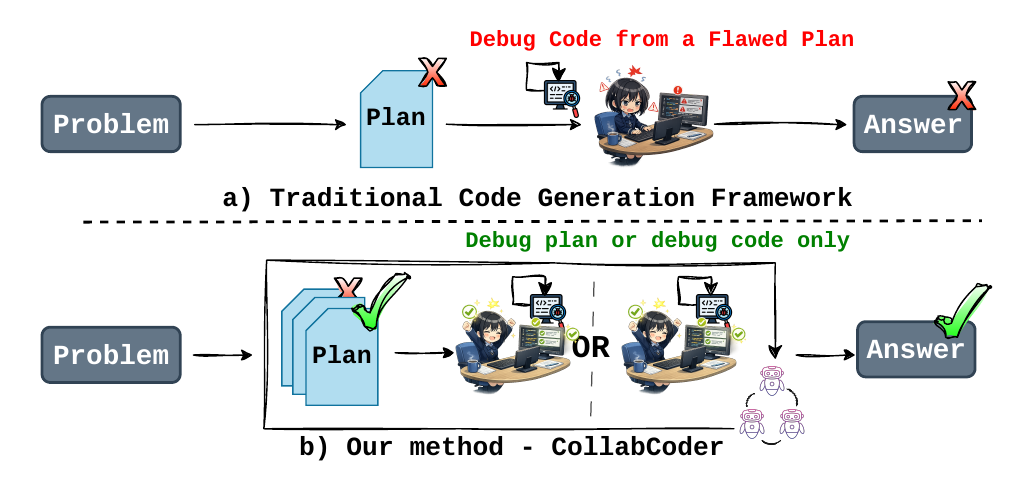}
    \caption{Overview of (a) a representative traditional code generation and (b) the proposed CollabCoder framework. Unlike conventional methods that rely on a fixed plan throughout code generation, CollabCoder allows the plan to be revised during execution. Multiple agents collaboratively assess intermediate outcomes and determine whether plan or code updates are required, enabling iterative refinement toward the final result.}
    \label{fig:intro}
\end{figure}
Code generation (known as program synthesis), a long-standing challenge in computer science, involves automatically generating programs from natural language requirements. The rapid growth of large language models (LLMs) has enabled fully executable code generation without human intervention~\cite{NijkampPHTWZSX23, LyuRRTT25, abs-2107-03374}. However, generating correct code for complex requirements poses challenges, particularly in advanced programming tasks~\cite{LiuXW023, DongJJL24}.

Early approaches to LLM-based code generation primarily relied on direct prompting, where models such as Codex~\cite{abs-2107-03374} generated code from natural language descriptions and input-output examples. More recent strategies have introduced structured prompting techniques like Chain-of-Thought reasoning~\cite{wei2022chain}, and retrieval-augmented generation~\cite{parvez2021retrieval}, which guide the model using similar prior problems and solutions. Despite these improvements, performance on complex code generation tasks remains limited, as generated outputs frequently fail to pass test cases and lack integrated bug-fixing mechanisms. To address this, the \textit{plan-before-code} paradigm~\cite{JiangDWFSLJJ24} has been proposed to separate high-level intent modeling from code synthesis. This idea has been incorporated into modern state-of-the-art systems, which typically adopt a dual-pass paradigm: In the first pass, a plan is generated and used to produce initial code with LLMs, while the second pass focuses on refining or debugging the resulting output (Figure~\ref{fig:intro}(a)). However, refinement is often superficial, driven by simple score-based retries that do not address root causes of failure. To overcome these limitations, recent works have proposed agent-based frameworks~\cite{IslamAP24, IslamAP25}, which decompose the generation process into modular components for retrieval, planning, and debugging in an iterative manner for improving the code through iterations. Despite their promise, these systems still suffer from two fundamental limitations. First, debugging remains largely reactive, with little support for contextual learning or explicit error attribution. Consequently, the system often produces repetitive and only marginally effective revisions, while failing to identify the root causes of errors or to leverage insights from prior unsuccessful attempts. This lack of principled error modeling not only undermines debugging efficiency, but also limits the system’s capacity to improve over successive iterations. Second, the planning module in existing approaches is typically held fixed throughout the debugging process, rather than being updated in response to code revisions and intermediate feedback. Such a static planning strategy prevents the planner and debugger from co-adapting over time, weakens coordination across different stages of the code generation pipeline, and further increases the complexity of repeatedly revising code under an already flawed plan. As a result, although these systems have shown encouraging potential, their inability to support adaptive debugging and iterative plan refinement remains a major barrier to robust and scalable performance.

Recognizing these limitations, we introduce CollabCoder, a novel Plan-Code Co-Evolution framework that addresses these challenges through multi-agent collaboration, adaptive feedback, and experience-driven learning, as illustrated in Figure~\ref{fig:intro}(b). Specifically, the main contributions of this study to agent-based code generation are threefold as follows:
(i) CollabCoder enables continuous improvement through a co-evolutionary process between planning and code generation. It leverages a multi-agent collaboration framework, termed Collaborative Decision-Making, in which planning, coding, and debugging agents work synergistically to decide whether to update the plan or refine the code at each iteration; (ii) To guide plan- or code-level updates, CollabCoder performs fine-grained analysis of multiple elements, including the plan, the generated code, and their alignment, rather than relying on superficial error-log inspection as in prior methods. These analyses are incorporated into a Reasoning Trajectory module, which integrates diagnostics from the current iteration with historical debugging strategies to iteratively improve both the plan and the code; and (iii) We conduct comprehensive evaluations of CollabCoder across a diverse set of benchmarks, ranging from simpler datasets (i.e., HumanEval and MBPP) to more challenging competitive programming benchmarks (i.e., xCodeEval and LiveCodeBench). The results demonstrate consistent improvements in terms of both accuracy and efficiency.
\section{Related Work}
\begin{figure*}[h]
  \centering
  \includegraphics[width=\linewidth]{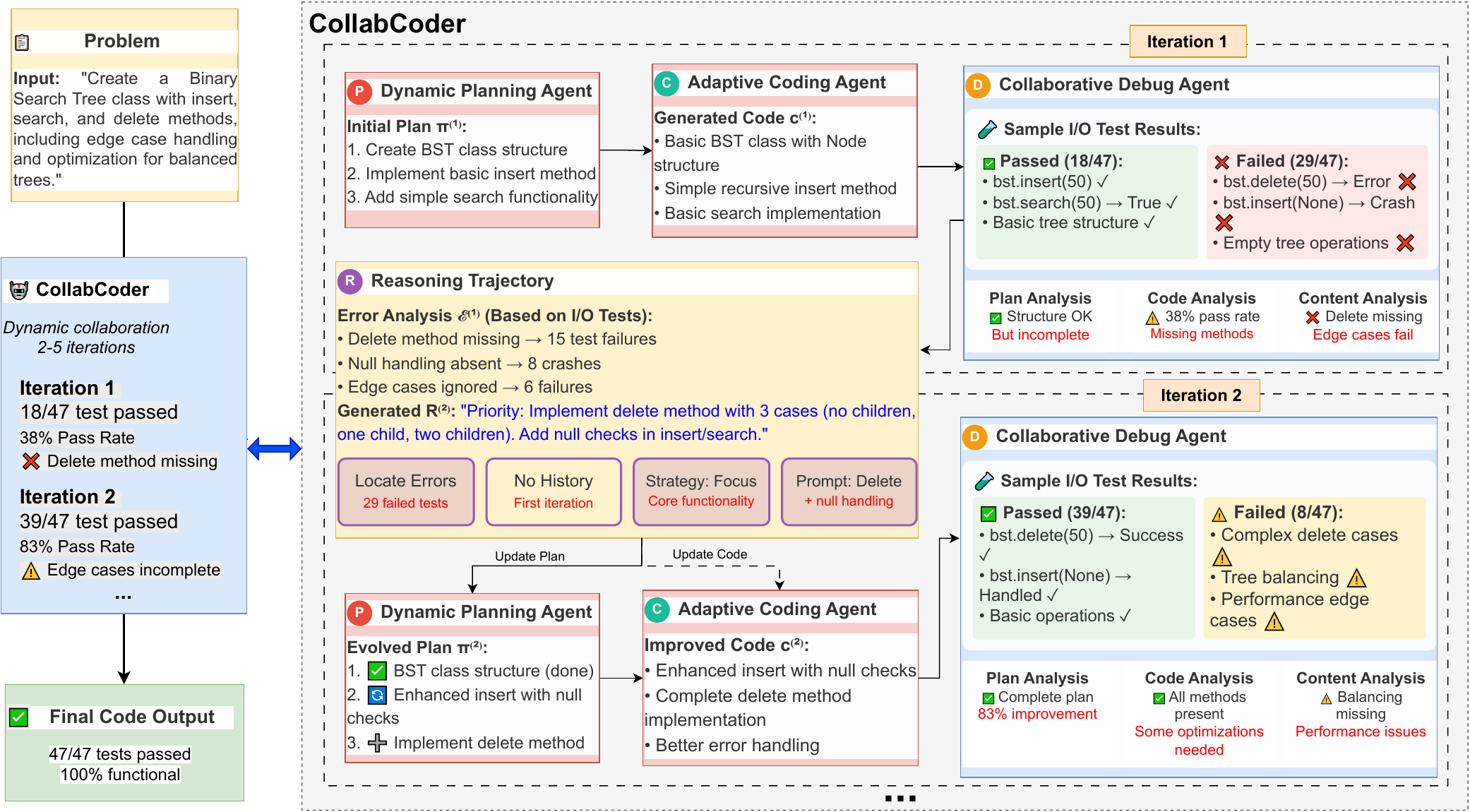}
  \caption{Architecture of the CollabCoder Framework. This diagram illustrates the co-evolutionary dynamics of the Dynamic Planning Agent and Adaptive Coding Agent, integrated with the Collaborative Debug Agent and Reasoning Trajectory Module, highlighting their continuous feedback loops and self-improving debugging mechanisms that enable adaptive plan evolution and code generation, overcoming the limitations of static planning in prior works.}
  \label{fig2}
\end{figure*}
\subsection{Code Generation Tasks}
Code generation has long been recognized as a fundamental challenge in software engineering~\cite{NiuL0022}. Traditional approaches primarily rely on training neural networks with task-specific annotated data methods, such as semantic parsing~\cite{RabinovichSK17} and retrieval-based techniques~\cite{ParvezACRC21}. Recently, several studies have leveraged pretrained language models (LMs) through continual training on both programming languages (PL) and natural languages (NL), enabling support for a variety of downstream NL-PL tasks, such as natural language code search and automatic code documentation generation~\cite{FengGTDFGS0LJZ20, 0034WJH21}. Nonetheless, the limited capacity of backbone language models restricts their utility in practical applications.

\subsection{LLM for Code Generation}
With their growing success, LLMs have demonstrated remarkable capabilities in code generation, driven by scaling model parameters to billions and training on extensive, diverse corpora with varied learning objectives. Recently, state-of-the-art closed-source models, such as GPT-4.1 and Claude 3.7 Sonnet, have emerged as powerful AI coding assistants. In parallel, open-source models like StarCoder~\cite{LiAZMKMMALCLZZW23}, Code Llama~\cite{abs-2308-12950}, DeepSeekCoder~\cite{abs-2401-14196}, and Qwen2.5-Coder~\cite{abs-2409-12186} have achieved significant breakthroughs, surpassing traditional methods across numerous benchmarks. Despite these advances, modern foundational LLMs for code still lack execution awareness, struggle to distinguish reasoning errors from implementation bugs, operate in a largely stateless manner without learning from past failures. As a result, developers must handle all testing and debugging themselves. 
\subsection{LLM Agents for Code Generation}
To overcome the limitations of standalone LLMs, recent work has proposed LLM agents that integrate planning, execution, and debugging into iterative workflows, such as MapCoder~\cite{IslamAP24}, CodeSIM~\cite{IslamAP25}, CodeAgent~\cite{ZhangLLSJ24}, ThinkCoder~\cite{ZhangLSCSY25}, and PairCoder~\cite{paircoder}. The core motivation behind these approaches is to enable LLMs to behave more like human developers within a digital environment by decomposing programming into multiple stages and iteratively refining solutions based on feedback. This design has led to more robust code generation than one-shot generation in many settings. While agent-based frameworks significantly improve robustness over single-pass generation, most existing approaches still follow a rigid trial-and-error paradigm. Their overall workflow is iterative, but not truly adaptive. Planning, coding, and debugging are executed in fixed sequences with limited adaptability, and execution feedback is primarily used to repair code rather than revise high-level reasoning. As a result, these systems lack explicit mechanisms to decide whether failures should be addressed at the plan or implementation level, often requiring extensive exploration across multiple plans or repeated code revisions.

\section{CollabCoder Framework}

\subsection{Overall Architecture}
\label{sec:overview}
As illustrated in the Figure \ref{fig2}, 
CollabCoder operates in an iterative, multi-agent manner and consists of three interacting agents, following the standard agentic paradigm commonly adopted in prior code generation systems~\cite{IslamAP24, IslamAP25}: A planning agent $\mathcal{A}_{\text{plan}}$, a coding agent $\mathcal{A}_{\text{code}}$, and a debugging agent $\mathcal{A}_{\text{debug}}$:
\begin{equation}
  \text{CollabCoder} = \langle \mathcal{A}_{\text{plan}}, \mathcal{A}_{\text{code}}, \mathcal{A}_{\text{debug}} \rangle.  
\end{equation}
In our framework, the debug agent is designed as a \textbf{Collaborative Decision-Making (CDM)} module, to determine whether the error should be addressed by updating the plan or the code, followed by a \textbf{Reasoning Trajectory (RT)} module to produce an updated debugging strategy in a learn-from-mistakes, self-improving manner:
\begin{equation}
  \mathcal{A}_{\text{debug}}= \langle
  \mathcal{A}_{\text{CDM}}, \mathcal{A}_{\text{RT}} 
  \rangle .
\end{equation}
This strategy guides plan or code refinement in the next iteration. The process repeats until the code satisfies all test cases or the iteration budget is exhausted. By combining CDM-driven adaptive decisions with RT-based self-improvement, CollabCoder enables dynamic co-evolution between planning and coding, avoiding the rigidity of static or fixed-planning approaches. 

\subsection{Methodology}

Given a coding problem defined by a natural-language problem description $P$, a set of coding templates $\mathcal{T}$ that guide the LLM toward generating code compatible with an evaluation oracle $\mathcal{O}$, and a collection of $Q$ test cases $\{(x_i, y_i)\}_{i=1}^Q$. At iteration $t$, CollabCoder maintains a solution plan $\pi^{(t)}$ and an executable program $c^{(t)}$. The program $c^{(t)}$ is executed on the test cases via $\mathcal{O}$ to produce observed outputs $\hat{y}_i$. All failing cases where $\hat{y}_i \neq y_i$ are aggregated into a test log $\mathcal{F}^{(t)}$.
\subsubsection{Collaborative Decision-Making}
\label{sec:cdm}
Based on the information maintained at iteration~$t$, the CDM module $\mathcal{A}_{\text{CDM}}$ operates in two main phases: (i) an analysis phase, where the system examines the current state from multiple complementary perspectives; and (ii) a decision phase, where some analyses are aggregated to determine whether to update the high-level plan or refine the generated code. During the analysis phase, $\mathcal{A}_{\text{CDM}}$ performs three complementary analyses, namely, plan-level analysis $\mathcal{E}_{\pi}^{(t)}$, code-level analysis $\mathcal{E}_{c}^{(t)}$, and plan-code alignment analysis $\mathcal{E}_{align}^{(t)}$:
\begin{equation}
\mathcal{E}_{\pi}^{(t)}, \mathcal{E}_{c}^{(t)}, \mathcal{E}_{align}^{(t)} 
= \mathcal{A}_{\text{CDM}}\bigl(P, \mathcal{F}^{(t)}, \pi^{(t)}, c^{(t)}\bigr), 
\label{CDM}
\end{equation}
where the \emph{plan-level analysis} $\mathcal{E}_{\pi}^{(t)}$ assesses whether the algorithmic reasoning encoded in $\pi^{(t)}$ is consistent with the observed failures, and identifies the underlying causes of plan-level errors. The \emph{code-level analysis} $\mathcal{E}_{c}^{(t)}$ focuses on diagnosing implementation errors in $c^{(t)}$ under the assumption that the plan itself is correct. The \emph{plan-code alignment analysis} $\mathcal{E}_{align}^{(t)}$ evaluates the semantic consistency between the plan and its realization in code, capturing cases where a correct plan fails due to incorrect or incomplete implementation. Sequentially, these analyses are jointly used to determine a decision
$D^{(t)} \in \mathcal{D} = \{0,1\}$, where $D^{(t)} = 0$ indicates updating the plan and $D^{(t)} = 1$ indicates updating the code. 

More concretely, the refinement decision at iteration $t$ is obtained by aggregating the three analysis signals
$\mathcal{E}_{\pi}^{(t)}$, $\mathcal{E}_{c}^{(t)}$, and $\mathcal{E}_{\mathrm{align}}^{(t)}$
through a consensus function $\mathcal{F}_{\mathrm{cons}}$.
This function is parameterized by a set of inter-module trust weights
$\mathcal{W}_{\mathrm{trust}} = \{ w_{\pi}, w_{c}, w_{\mathrm{align}} \}$,
which are fixed hyperparameters shared across tasks to ensure stable and consistent decision behavior.
Each weight $w_i \geq 0$ reflects the relative reliability of the corresponding analysis module
$i \in \mathcal{H} = \{ \pi, c, \mathrm{align} \}$,
with the normalization constraint $\sum_i w_i = 1$. The resulting collaborative decision $D^{(t)} \in \mathcal{D}$ is given by:
\begin{equation}
D^{(t)} = \mathcal{F}_{\mathrm{cons}}\!\left(
\mathcal{E}_{\pi}^{(t)}, \mathcal{E}_{c}^{(t)}, \mathcal{E}_{align}^{(t)}, \mathcal{W}_{\mathrm{trust}}
\right).
\end{equation}

Specifically, the decision at iteration $t$ is obtained by maximizing an aggregated score over the decision space $\mathcal{D}$, which jointly accounts for individual analysis confidence and cross-analysis consistency:
\begin{equation}
\begin{aligned}
D^{(t)} = \arg\max_{d \in \mathcal{D}}
\sum_{i \in \mathcal{H}}
w_i \cdot \varphi_{i,d}^{(t)} \cdot \phi_{\mathcal{H} \setminus \{i\},d}^{(t)} .
\end{aligned}
\label{eq:cdm}
\end{equation}

Here, $\varphi_{i,d}^{(t)} \in [0,1]$ denotes the confidence score indicating how strongly analysis $i$ supports decision $d$, while $\phi_{\mathcal{H} \setminus \{i\},d}^{(t)} \in [0,1]$ measures the consistency of decision $d$ with the remaining analyses. All confidence and consistency scores $\{ \varphi_{i,d}^{(t)},\phi_{\mathcal{H} \setminus \{i\},d}^{(t)} \}$ are jointly produced in a single LLM invocation, conditioned on the set of analyses $\{\mathcal{E}_\pi^{(t)}, \mathcal{E}_c^{(t)}, \mathcal{E}_{\text{align}}^{(t)}\}$.

\subsubsection{Reasoning Trajectory Module}
\label{sec:rt}
The RT module enables iterative self-improvement by maintaining a persistent debugging strategy across iterations. Unlike prior approaches that treat each failure independently, RT explicitly accumulates historical diagnostic information and leverages it to guide subsequent refinements in a learn-from-mistakes manner. At iteration $t$, RT maintains a reasoning state $R^{(t)}$ that summarizes prior debugging insights and refinement patterns. This state is updated by jointly considering historical context and current diagnostic signals. Formally, the update rule is defined as:
\begin{equation}
R^{(t)} = \mathcal{A}_{\mathrm{RT}}\!\left(
R^{(t-1)}, \mathcal{E}_{X}^{(t)}, P, X^{(t)}, \mathcal{F}^{(t)}
\right),
\label{eq:rt}
\end{equation}
where we define $X^{(t)}$ as a unified refinement state to denote the solution component selected for refinement at iteration $t$ and $\mathcal{E}_{X}^{(t)}$ denotes the diagnostic analysis corresponding to the current refinement target $X^{(t)}$. Specifically,
\begin{equation}
   X^{(t)} \triangleq \mathbb{I}[D^{(t)}=0]\cdot \pi^{(t)} + \mathbb{I}[D^{(t)}=1]\cdot c^{(t)} . 
\end{equation}
The updated reasoning strategy $R^{(t)}$ is subsequently used to guide the refinement operator in the next iteration, influencing how the selected plan or code component is revised. Concretely, the next refinement state is obtained by applying the decision-conditioned refinement operator to the current state and the updated debugging strategy:
\begin{equation}
X^{(t+1)} = \mathcal{A}_X\!\left(X^{(t)}, P, R^{(t)}, \mathcal{T}^{(t)}, \mathcal{F}^{(t)}\right),    
\end{equation}
where $\mathcal{A}_X \in \{\mathcal{A}_{\mathrm{plan}}, \mathcal{A}_{\mathrm{code}}\}$ is selected according to the collaborative decision $D^{(t)}$. The coding template $\mathcal{T}^{(t)}$ is provided only when code refinement is selected, i.e.,
$\mathcal{T}^{(t)} \triangleq \mathbb{I}[D^{(t)}=1]\cdot \mathcal{T}$, ensuring that structural constraints are enforced during code updates while plan refinement remains template-independent. 

Technically, by jointly conditioning on multiple sources of information, namely, historical debugging strategies captured in $R^{(t-1)}$, localized error diagnoses from the current iteration $\mathcal{E}_{X}^{(t)}$, the original problem specification $P$, the current plan or code $X^{(t)}$, and concrete failure evidence from $\mathcal{F}^{(t)}$, the RT module produces a structured reasoning strategy that highlights recurring error patterns, refines corrective heuristics, and avoids repeating ineffective fixes observed in previous iterations. Moreover, by explicitly modeling debugging as a stateful reasoning process, RT improves convergence stability and reduces redundant trial-and-error behaviors commonly observed in stateless debugging approaches.


\section{Experiment}
\subsection{Datasets}
We evaluate CollabCoder on two widely used benchmark datasets for code generation, HumanEval~(HE)~\cite{abs-2107-03374} and MBPP~\cite{abs-2108-07732}, along with their extended versions enriched with additional test cases, HumanEval-ET (HE-ET) and MBPP-ET, respectively~\cite{DongDJLLJ25}. For evaluating performance on complex, contest-level problems, we utilize LiveCodeBench~\cite{JainHGLYZWSSS25} and xCodeEval~\cite{khan-etal-2024-xcodeeval}, recently established benchmarks for assessing LLM-based code generation. 


\subsection{Baselines and Experimental Setting}
We compare CollabCoder with direct LLM-based baselines, including Chain-of-Thought (CoT) prompting~\cite{YangZCZZC24} and Self-Planning~\cite{JiangDWFSLJJ24}, as well as recent agent-based frameworks such as MapCoder~\cite{IslamAP24}, CodeSIM~\cite{IslamAP25}, and ThinkCoder~\cite{ZhangLSCSY25}. All methods are evaluated under identical backbone LLM settings, covering both proprietary models (GPT-4o mini) and open-source models (Seed-Coder-8B and Qwen2.5-Coder-32B). We report effectiveness in terms of zero-shot Pass@1 accuracy on HumanEval, MBPP, and their extended variants (HumanEvalET and MBPP-ET), along with efficiency metrics including average token consumption and the number of API calls per problem. For agent-based baselines, including MapCoder, CodeSIM, and ThinkCoder, we use a fixed exploration budget of $k=5$ planning iterations and a refinement budget of $n=5$ debugging iterations for MapCoder and CodeSIM, while setting $k=1$ and $n=20$ for ThinkCoder, following the best configurations reported in their respective papers. For CollabCoder, we use $t=5$ iterations, matching the same budget convention adopted by the baselines with $k=1$ and $n=5$. We fix the trust weights of CollabCoder’s Collaborative Decision-Making module to $w_{\pi}=0.4$, $w_{c}=0.3$, and $w_{\text{align}}=0.3$.

\subsection{Main Results}
\begin{table*}[ht]
\centering
\small
\begin{tabular}{lccccc|cccc}
\toprule
\multicolumn{6}{c}{\textbf{Accuracy} $\uparrow$} & \multicolumn{4}{c}{\textbf{Efficiency} $\downarrow$} \\
\toprule
\textbf{Method} & \textbf{HE} & \textbf{HE-ET} &  \textbf{MBPP} & \textbf{MBPP-ET} & \textbf{Average} & \textbf{k} & \textbf{n} & \textbf{Token I/O} & \textbf{API calls}\\
\midrule

\rowcolor{headergray}
\multicolumn{10}{l}{\textbf{Backbone LLM: Seed-Coder-8B}}	\\
Direct &	18.90	& 17.07  & 59.19	 & 38.03		&  33.30 & -- & -- & \phantom{0}175.27 / \phantom{0}289.66 & 1.00 \\
CoT &	82.32	& 75.00  &	75.06 & 50.13		& 70.63 & -- & -- & 1087.79 / \phantom{0}182.70 & 1.00 \\
Self-Planning &	82.32	& 71.34  &	74.06 & 51.13 & 69.71 & -- & -- & 2154.48 / 1018.15 & 2.00 \\
MapCoder & 79.88 & 70.12 & 73.55 & 49.12 & 68.78 &5 & 5 & 6323.28 / 3022.59 & 9.84 \\
CodeSIM & \textbf{90.24} & \underline{76.20} & \underline{82.00} & \underline{53.65} & \underline{75.51} & 5 & 5 & 4154.30 / 3943.08 & 6.69 \\
ThinkCoder & 82.32 & 73.78 & 76.83 & 51.39 & 71.08 & 1 & 20 & 1613.02 / 1544.64 & 4.56 \\
\textbf{CollabCoder (Ours)} & \underline{87.20} & \textbf{78.05} & \textbf{83.37} & \textbf{56.42} & \textbf{76.26} & 1 & 5 & \textbf{4219.78 / 1964.03} & \textbf{5.06} \\

\midrule
\rowcolor{headergray}
\multicolumn{10}{l}{\textbf{Backbone LLM: Qwen2.5-Coder-32B}}\\
Direct & 85.37 & 75.61 & 79.09 & 54.91 & 75.00 & -- & -- & \phantom{0}128.93 / \phantom{0}365.61 & 1.00 \\
CoT & 90.24 & \underline{81.70} & 83.38 & \underline{59.44} & 79.15 & -- & -- & \phantom{0}948.07 / \phantom{00}89.79  & 1.00 \\
Self-Planning & 87.80 & 76.83 & 77.33 & 53.15 & 74.70 & -- & -- & 1635.33 / 0640.73 & 2.00 \\

MapCoder & 90.24  & 79.00  & 86.80 & \textbf{59.95} & 79.84 &5 & 5 
& 5848.39 / 3309.55 & 9.05 \\

CodeSIM & \underline{93.29} & \underline{81.70} & \underline{87.20}& 58.70 & \underline{80.22} & 5 & 5 & 2191.03 / 2593.04 & 4.87 \\

ThinkCoder & 88.41 & 79.88 & 85.89 & 53.90 & 77.02 
& 1 & 20 & 1128.67 / 1404.25 & 2.99 \\

\textbf{CollabCoder (Ours)} & \textbf{95.73} & \textbf{84.15} & \textbf{90.17} & \textbf{59.95} & \textbf{82.50} 
& 1 & 5 & \textbf{2468.22 / 1606.88} & \textbf{4.12} \\

\midrule
\rowcolor{headergray}
\multicolumn{10}{l}{\textbf{Backbone LLM: GPT-4o mini}}\\
Direct         & 85.97 & 76.22 & 75.82 & 52.14 & 72.54 & -- & -- & \phantom{0}105.98 / \phantom{0}396.46  & 1.00 \\
CoT            & 85.97 & 78.66 & 78.59 & 54.66 & 74.47 & -- & -- & \phantom{0}946.50 / \phantom{0}129.60 & 1.00 \\
Self-Planning  & 82.32 & 74.39 & 78.84 & 53.65 & 72.30 & -- & -- & 1716.58 / \phantom{0}854.74 & 2.00 \\

MapCoder       & 90.24 & 79.88 & 84.13 & 56.93 & 77.80 &5 & 5 & 5767.54 / 2965.21 & 10.10 \\
CodeSIM        & \underline{94.51} & \underline{81.70} & \underline{89.92} & \underline{59.95} & \underline{81.52}
               & 5 & 5 & 2397.89 / 2688.32 & 5.16 \\
ThinkCoder     & 90.85 & \underline{81.70} & 81.61 & 55.92 & 77.52 & 1 & 20 & 1007.58 / 1172.93  & 2.20 \\
\textbf{CollabCoder (Ours)} 
               & \textbf{96.34} & \textbf{84.76} & \textbf{91.69} & \textbf{60.20} & \textbf{83.25} 
               & 1 
               & 5
               & \textbf{2993.33 / 1781.21} & \textbf{5.06} \\
\bottomrule
\end{tabular}
\caption{Accuracy and efficiency comparison across multiple benchmark datasets and backbone LLMs.
Accuracy is measured by Pass@1, while efficiency is assessed using Token In/Out and the number of API calls.
Token In/Out is averaged across all API calls and datasets, with its detailed formulation provided in Appendix~\ref{app:metric}. In the \textit{Accuracy} section, boldface indicates the best-performing method for each backbone, and underlined values denote the second-best. In the \textit{Efficiency} section, boldface highlights the results of our method.}
\label{tab:main}
\end{table*}

\begin{table*}[ht]
\centering
\begin{adjustbox}{max width=\linewidth}
\begin{tabular}{l|c|c|c|c|c}
\hline
\textbf{Method} 
& \textbf{LiveCodeBench} 
& \textbf{xCodeEval} 
& \textbf{Average}
& \textbf{Token I/O}
& \textbf{API Calls} \\
\hline
MapCoder 
& 34.82 & 40.57 & 37.70 
& 28437.65 / 17692.18 & 22.41 \\

CodeSIM 
& 36.61 & 42.45 & 39.53
& 20907.82 / 13151.10 & 17.16	 \\


\textbf{CollabCoder (Ours)} 
& \textbf{41.96} & \textbf{47.16} & \textbf{44.56} 
& \textbf{15155.93 / \phantom{0}4491.37} & \textbf{12.27} \\
\hline
\end{tabular}
\end{adjustbox}
\caption{Pass@1 accuracy and efficiency comparison on contest-level code generation benchmarks. Token I/O denotes the average number of input and output tokens per problem, and API Calls indicate the average number of model invocations. All methods are evaluated using GPT-4o mini.}
\label{tab:competitive}
\end{table*}
\subsubsection{Performance on basic code generation}
The results in Table~\ref{tab:main} highlight a clear trade-off between effectiveness and efficiency among different approaches. Lightweight methods such as Direct Prompting, CoT, Self-Planning, and ThinkCoder generally consume fewer LLM resources but yield limited performance gains. For instance, on Seed-Coder-8B, Direct Prompting achieves an average accuracy of only 33.30 despite minimal token usage (175.27 / 289.66), while ThinkCoder reaches 71.08 with moderate efficiency cost. More agentic frameworks, such as MapCoder and CodeSIM, substantially improve accuracy by incorporating multi-stage planning, simulation, and iterative debugging; however, these gains come at the cost of significantly higher computational overhead. For instance, on Seed-Coder-8B, MapCoder consumes 6323.28 / 3022.59 tokens with 9.84 API calls per problem, while CodeSIM requires 4154.30 / 3943.08 tokens and 6.69 API calls to achieve an average accuracy of 75.51. This overhead can be attributed to their effective inference complexity, which grows on the order of $O(nk)$. When an initial plan is misaligned, these methods may repeatedly expend API calls debugging code derived from fundamentally flawed plans, leading to substantial redundant computation.

To address this limitation, CollabCoder is designed to preserve the strengths of agentic reasoning while reducing unnecessary computation. By adopting a plan-code co-evolution strategy, CollabCoder jointly refines high-level plans and low-level code within a single evolving trajectory, thereby reducing the effective inference complexity to depend only on the number of refinement iterations. As a result, CollabCoder achieves comparable or superior accuracy with substantially lower resource consumption. Across Seed-Coder-8B, Qwen2.5-Coder-32B, and GPT-4o mini, it reduces total token usage by approximately 30-50\% compared to MapCoder and by 10-25\% compared to CodeSIM, while consistently attaining higher average accuracy. Overall, these results demonstrate that CollabCoder provides a more favorable balance between effectiveness and efficiency than existing baselines.
\subsubsection{Complex Code Generation}
We further evaluate the proposed method on complex, contest-level code generation tasks to assess its capability in solving programming problems that closely resemble real-world competitive coding scenarios. To this end, we adopt two widely used benchmarks, LiveCodeBench and xCodeEval, which are specifically designed to evaluate the robustness of code generation systems across different difficulty levels. In this setting, we compare CollabCoder against MapCoder and CodeSIM, two state-of-the-art agentic approaches for complex competitive programming. 

For fairness and consistency, all methods are evaluated using GPT-4o mini. We exclude Seed-Coder-8B and Qwen2.5-Coder-32B from this evaluation, as preliminary experiments indicate that their limited model capacity leads to uniformly low performance on contest-level benchmarks, thereby obscuring meaningful performance differences. Table~\ref{tab:competitive} summarizes the Pass@1 accuracy and efficiency of different methods on these competitive programming benchmarks.
\begin{figure*}[h]
    \centering
    \includegraphics[width=0.975\linewidth]{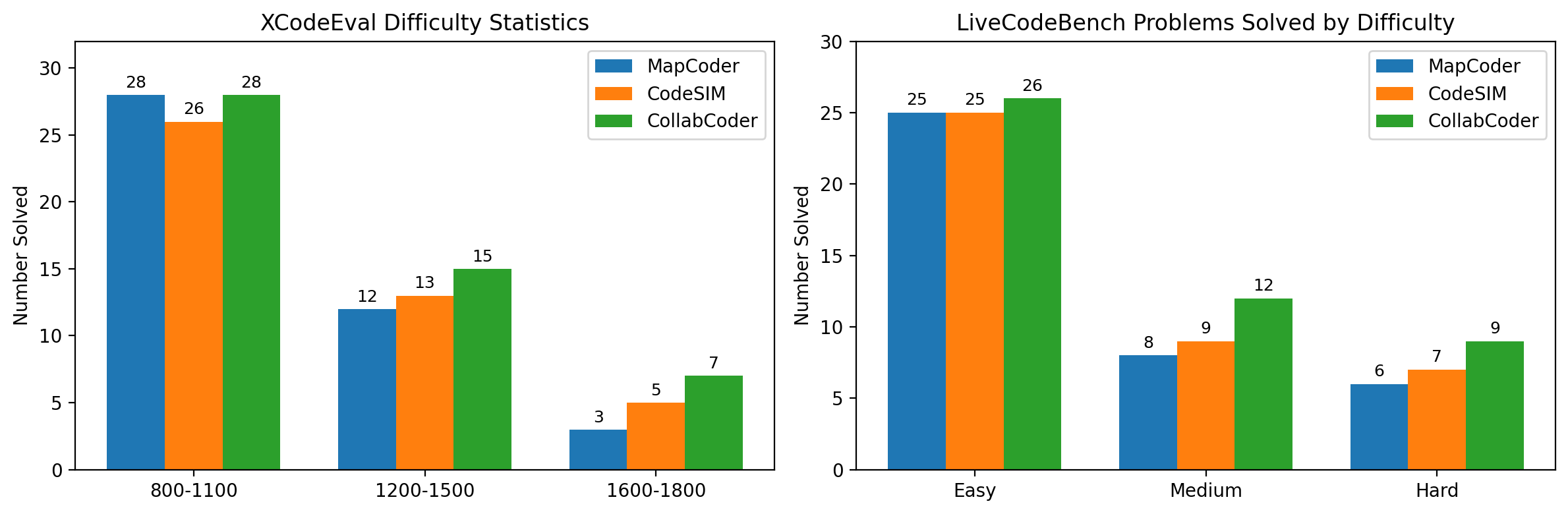}
    \caption{Distribution of solved competitive programming problems across different difficulty levels, illustrating the proportion of problems completed at each level.}
    \label{fig:difficulty}
\end{figure*}
Accordingly, compared to basic code generation settings, the advantages of CollabCoder become more pronounced on complex, contest-level code generation tasks. As shown in Table~\ref{tab:competitive}, CollabCoder achieves a Pass@1 accuracy of 41.96\% on LiveCodeBench and 47.16\% on xCodeEval, outperforming MapCoder by approximately 6.6-7.1 percentage points and CodeSIM by 4.7-5.3 points across the two benchmarks. In addition to accuracy gains, CollabCoder exhibits a clear advantage in computational efficiency, reducing total token consumption by approximately 57\% compared to MapCoder and 42\% compared to CodeSIM. These consistent improvements demonstrate CollabCoder’s strong effectiveness and efficiency on especially challenging tasks. 

Furthermore, to gain a finer-grained perspective on model performance, we further examine the distribution of solved problems across difficulty levels (Figure~\ref{fig:difficulty}). 
On xCodeEval, CollabCoder achieves performance comparable to MapCoder in the easiest difficulty range (800–1100), with both methods solving 28 problems, while CodeSIM solves 26. As task difficulty increases, CollabCoder tends to maintain more stable performance. In the 1200–1500 range, CollabCoder solves 15 problems, compared to 12 for MapCoder and 13 for CodeSIM. In the hardest range (1600–1800), CollabCoder solves 7 problems, whereas MapCoder and CodeSIM solve 3 and 5 problems, respectively. These results suggest that CollabCoder experiences a milder performance degradation as problem difficulty increases, with the differences becoming more noticeable in the medium and hard difficulty ranges. A similar trend is observed on LiveCodeBench. For medium-difficulty problems, CollabCoder solves 12 tasks, compared to 8 for MapCoder and 9 for CodeSIM. On hard problems, CollabCoder remains competitive with 9 solved tasks, slightly higher than MapCoder (6) and CodeSIM (7). Overall, these results indicate that CollabCoder maintains relatively stable effectiveness across different difficulty levels and demonstrates better adaptability when handling more complex, contest-level programming tasks.
\subsection{Self-improving Debugging Analysis}
\label{sec:self_improving_analysis}
\begin{figure}[ht]
    \centering
    \includegraphics[width=\linewidth]{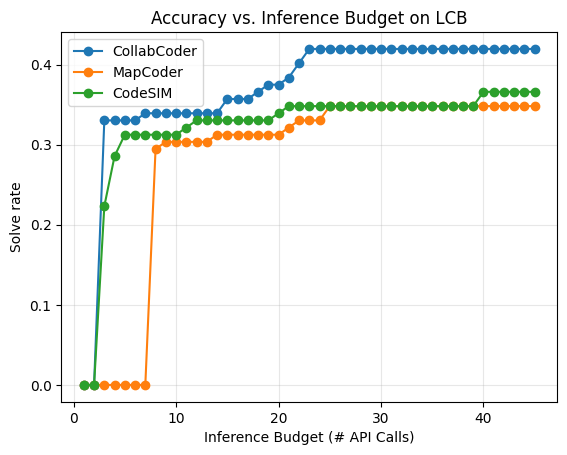}
    \caption{Accuracy vs.\ Inference budget on LiveCodeBench benchmark.}
    \label{fig:accuracy_budget}
\end{figure}

Figure~\ref{fig:accuracy_budget} illustrates the accuracy--budget trade-off of different debugging strategies on the LiveCodeBench benchmark. Analogous results on xCodeEval are provided in Appendix~\ref{sec:additional_self_improve_xcodeeval}. Examining these trade-offs on competitive code generation benchmarks, where iterative debugging plays a critical role, provides further insight into how each method improves code quality under constraints on the number of API calls.
On LiveCodeBench, CollabCoder demonstrates a clear advantage in the low-budget regime. With an inference budget of only 10 API calls, CollabCoder achieves a solve rate of 33.93\%, outperforming both MapCoder (30.36\%) and CodeSIM (31.25\%). This result indicates that, under identical budget constraints, CollabCoder more effectively integrates high-level reasoning signals through accumulated experience and structured analysis in its Reasoning Trajectory module. As a result, it is able to translate limited feedback into targeted and meaningful improvements, rather than expending inference budget on largely trial-and-error debugging. In contrast, MapCoder and CodeSIM exhibit less efficient budget utilization, requiring additional API calls to achieve comparable accuracy gains.
\subsection{Collaborative Decision-Making Analysis}
\begin{figure}[h]
    \centering
    \includegraphics[width=0.955\linewidth]{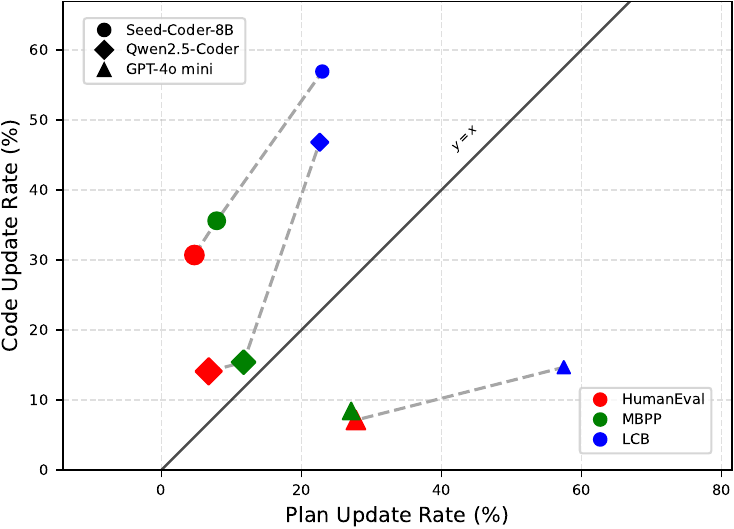}
    \caption{Relationship between \textit{Plan Update Rate} and \textit{Code Update Rate} across different backbone LLMs and datasets. Marker shapes denote backbone models, colors indicate datasets, and marker sizes are proportional to accuracy on the corresponding dataset. The update rate is a normalized metric that allows comparison across different datasets, defined as the ratio of the number of updates to the total number of iterations; further details are provided in Appendix~\ref{app:metric}.
}
    \label{fig:cdm_analysis_rate}
\end{figure}

To better understand the decision-making behavior of the CDM module, we analyze its behavior across different backbone models and datasets with varying levels of difficulty. An intuitive visualization is provided in Figure~\ref{fig:cdm_analysis_rate} and more detailed statistics are reported in Table~\ref{tab:cdm_analysis} (Appendix~\ref{CDM_analysis}).
\paragraph{Effect of Dataset Difficulty.} Across all backbones, the frequency of CDM-triggered updates consistently increases as dataset difficulty rises. As a result, a common trend across all three backbones is that the update rate increases (at both the code and plan levels) as the datasets become more challenging. This monotonic trend indicates that harder benchmarks introduce a higher incidence of failure cases, thereby requiring CDM to intervene more frequently. Such behavior suggests that CDM effectively adapts its intervention rate to task complexity, providing additional corrective signals when the problem space becomes harder to navigate.

\paragraph{Backbone-Specific Update Dynamics.} Despite the shared trend described above, the allocation between plan-level and code-level updates differs markedly across backbones. Code-specialized models predominantly rely on code-level revisions, with plan update rates approximately two to three times lower than code update rates. In contrast, the general-purpose GPT-4o mini allocates a substantially larger fraction of its CDM interventions to plan-level revisions. As shown in Figure~\ref{fig:cdm_analysis_rate}, the diagonal line $y = x$ clearly separates data points corresponding to these two classes of models.

\paragraph{Optimization Perspective.} We interpret this divergence from an optimization perspective. Code-specialized models tend to operate within a narrow neighborhood of the initial plan, applying incremental code-level modifications. When the initial plan is suboptimal, such localized adjustments are prone to converging to local minima, where repeated implementation-level fixes fail to correct a flawed high-level strategy. In contrast, strong general-purpose models demonstrate a higher sensitivity to diagnosing failures that originate from inappropriate high-level strategies rather than isolated coding errors. In these cases, CDM more frequently selects plan-level updates, enabling the model to escape local minima by exploring alternative regions of the solution space.
\begin{table*}[t]
\centering
\resizebox{\textwidth}{!}{%
\begin{tabular}{lccccccc}
\toprule
\textbf{Model} & \textbf{w/ CDM} & \textbf{w/ RT} & \textbf{HumanEval} & \textbf{HumanEvalET} & \textbf{MBPP} & \textbf{MBPP-ET} & \textbf{Avg} \\
\midrule
Seed-Coder-8B & $\times$ & $\checkmark$ & 85.37\small{{\textcolor{red!70}{$\downarrow$1.83}}} & 75.00\small{{\textcolor{red!70}{$\downarrow$3.05}}} & 76.07\small{{\textcolor{red!70}{$\downarrow$7.30}}} & 51.64\small{{\textcolor{red!70}{$\downarrow$4.78}}} & 72.02\small{{\textcolor{red!70}{$\downarrow$4.24}}} \\
Seed-Coder-8B & $\checkmark$ & $\times$ & 85.00\small{{\textcolor{red!70}{$\downarrow$2.20}}} & 75.61\small{{\textcolor{red!70}{$\downarrow$2.44}}} & 79.34\small{{\textcolor{red!70}{$\downarrow$4.03}}} & 51.64\small{{\textcolor{red!70}{$\downarrow$4.78}}} & 72.90\small{{\textcolor{red!70}{$\downarrow$3.36}}} \\
Seed-Coder-8B & $\checkmark$ & $\checkmark$ & \textbf{87.20} & \textbf{78.05} & \textbf{83.37} & \textbf{56.42} & \textbf{76.26} \\
\midrule
Qwen2.5-Coder-32B & $\times$ & $\checkmark$ & 90.24\small{{\textcolor{red!70}{$\downarrow$5.49}}} & 81.10\small{{\textcolor{red!70}{$\downarrow$3.05}}} & 83.38\small{{\textcolor{red!70}{$\downarrow$6.79}}} & 56.93\small{{\textcolor{red!70}{$\downarrow$3.02}}} & 77.91\small{{\textcolor{red!70}{$\downarrow$4.59}}} \\
Qwen2.5-Coder-32B & $\checkmark$ & $\times$ & 92.68\small{{\textcolor{red!70}{$\downarrow$3.05}}} & 82.92\small{{\textcolor{red!70}{$\downarrow$1.23}}} & 88.41\small{{\textcolor{red!70}{$\downarrow$1.76}}} & 58.94\small{{\textcolor{red!70}{$\downarrow$1.01}}} & 80.74\small{{\textcolor{red!70}{$\downarrow$1.76}}} \\
Qwen2.5-Coder-32B & $\checkmark$ & $\checkmark$ & \textbf{95.73} & \textbf{84.15} & \textbf{90.17} & \textbf{59.95} & \textbf{82.50} \\
\bottomrule
\end{tabular}
}
\caption{Ablation study on the impact of CDM and RT. We evaluate different variants of CollabCoder with and without CDM and RT on the HumanEval and MBPP benchmarks using the Pass@1 metric. \textit{w/ CDM} and \textit{w/ RT} indicate that the corresponding component is enabled in the pipeline.}
\label{tab:ablation}
\end{table*}
\subsection{Ablation Study}
To better understand the contribution of each component in CollabCoder, we conduct an ablation study on its two core modules: Collaborative Decision-Making (CDM) (Equation~\ref{eq:cdm}) and Reasoning Trajectory (RT) (Equation~\ref{eq:rt}). For CDM, we replace it with a conventional debugging mechanism that always updates the code, following prior work~\cite{IslamAP24, IslamAP25}. For RT, we remove it and directly use the output of CDM to guide the next iteration. The results in Table~\ref{tab:ablation} highlight the crucial role of both modules and confirm their importance to the overall performance of CollabCoder. First, replacing CDM with a standard debugging approach causes a significant drop in performance across all datasets for both base models. This suggests that CDM is essential for coordinating decisions during iterative refinement, enabling more robust code updates than conventional debugging strategies. Second, removing RT also leads to a consistent decline, although the drop is generally smaller than that caused by removing CDM. This indicates that while CDM provides the foundation for collaborative improvement, RT further improves solution quality by guiding the model along structured reasoning paths instead of relying only on raw outputs. Notably, the full version of CollabCoder achieves the best average performance on both base models, showing that the two modules are complementary. For the larger Qwen2.5-Coder-32B, the benefits are even more pronounced, with the full model outperforming both ablated variants by a clear margin, especially on HumanEval and MBPP. This suggests that CollabCoder scales well with stronger base models and better exploits their capacity when both CDM and RT are used together.

\section{Conclusion and Future Work}
This study presented CollabCoder, a collaborative multi-agent framework that addresses fundamental limitations in existing code generation systems by introducing dynamic planning, adaptive coding, and self-improving debugging. Unlike conventional static pipelines, CollabCoder enables continuous co-evolution between plans and code through a collaborative analysis engine that synthesizes insights across planning, coding, and debugging. Experimental results across standard, extended, and contest-level benchmarks confirm that CollabCoder achieves higher accuracy, robustness, and adaptability, while maintaining computational efficiency. The framework establishes a new paradigm for automated programming, where coordinated agent collaboration surpasses isolated strategies.
For future work, we plan to extend CollabCoder by integrating formal verification techniques, exploring multi-modal programming tasks, and enhancing semantic alignment between evolving plans and specifications. These directions position CollabCoder as a promising foundation for next-generation collaborative AI systems in software development.
\section*{Limitations}
While CollabCoder demonstrates strong empirical performance across benchmarks, it is not without limitations. Firstly, our approach remains heavily dependent on the capability of the underlying LLM backbone. In particular, the analysis and collaborative decision-making stages require strong reasoning and code understanding, which may limit the effectiveness of CollabCoder when using weaker or resource-constrained models. Secondly, CollabCoder relies on a fixed and limited number of sample I/O pairs for debugging. While helpful for iterative refinement, their limited coverage may restrict robustness against diverse edge cases. Finally, future work will focus on improving automatic additional test case generation to enhance sample I/O quality, as well as adapting CollabCoder to lighter-weight backbone models to improve efficiency and accessibility.


\bibliography{custom}

@inproceedings{NijkampPHTWZSX23,
  author       = {Erik Nijkamp and
                  Bo Pang and
                  Hiroaki Hayashi and
                  Lifu Tu and
                  Huan Wang and
                  Yingbo Zhou and
                  Silvio Savarese and
                  Caiming Xiong},
  title        = {CodeGen: An Open Large Language Model for Code with Multi-Turn Program
                  Synthesis},
  booktitle    = {The Eleventh International Conference on Learning Representations,
                  {ICLR} 2023, Kigali, Rwanda, May 1-5, 2023},
  publisher    = {OpenReview.net},
  year         = {2023},
  bibsource    = {dblp computer science bibliography, https://dblp.org}
}

@article{LyuRRTT25,
  author    = {Michael R. Lyu and Baishakhi Ray and Abhik Roychoudhury and Shin Hwei Tan},
  title     = {Automatic Programming: Large Language Models and Beyond},
  journal   = {ACM Trans. Softw. Eng. Methodol.},
  volume    = {34},
  number    = {6},
  pages     = {140:1--140:33},
  year      = {2025},
  publisher = {Association for Computing Machinery},
  doi       = {10.1145/3708519},
  url       = {https://doi.org/10.1145/3708519}
}

@inproceedings{wei2022chain,
  title     = {Chain-of-Thought Prompting Elicits Reasoning in Large Language Models},
  author    = {Jason Wei and Xuezhi Wang and Dale Schuurmans and Maarten Bosma and Ed H. Chi and Quoc V. Le and Denny Zhou},
  booktitle = {Advances in Neural Information Processing Systems (NeurIPS)},
  year      = {2022},
  url       = {https://proceedings.neurips.cc/paper_files/paper/2022/file/9d5609613524ecf4f15af0f7b31abca4-Paper-Conference.pdf}
}

@inproceedings{parvez2021retrieval,
  title     = {Retrieval Augmented Code Generation and Summarization},
  author    = {Parvez, Md Rizwan and Ahmad, Wasi and Chakraborty, Saikat and Ray, Baishakhi and Chang, Kai-Wei},
  booktitle = {Findings of the Association for Computational Linguistics: EMNLP 2021},
  pages     = {2719--2734},
  year      = {2021},
  month     = nov,
  address   = {Punta Cana, Dominican Republic},
  publisher = {Association for Computational Linguistics},
  doi       = {10.18653/v1/2021.findings-emnlp.232},
  url       = {https://aclanthology.org/2021.findings-emnlp.232/}
}

@article{abs-2107-03374,
  author       = {Mark Chen and
                  Jerry Tworek and
                  Heewoo Jun and
                  Qiming Yuan and
                  Henrique Pond{\'{e}} de Oliveira Pinto and
                  Jared Kaplan and
                  Harri Edwards and
                  Yuri Burda and
                  Nicholas Joseph and
                  Greg Brockman and
                  Alex Ray and
                  Raul Puri and
                  Gretchen Krueger and
                  Michael Petrov and
                  Heidy Khlaaf and
                  Girish Sastry and
                  Pamela Mishkin and
                  Brooke Chan and
                  Scott Gray and
                  Nick Ryder and
                  Mikhail Pavlov and
                  Alethea Power and
                  Lukasz Kaiser and
                  Mohammad Bavarian and
                  Clemens Winter and
                  Philippe Tillet and
                  Felipe Petroski Such and
                  Dave Cummings and
                  Matthias Plappert and
                  Fotios Chantzis and
                  Elizabeth Barnes and
                  Ariel Herbert{-}Voss and
                  William Hebgen Guss and
                  Alex Nichol and
                  Alex Paino and
                  Nikolas Tezak and
                  Jie Tang and
                  Igor Babuschkin and
                  Suchir Balaji and
                  Shantanu Jain and
                  William Saunders and
                  Christopher Hesse and
                  Andrew N. Carr and
                  Jan Leike and
                  Joshua Achiam and
                  Vedant Misra and
                  Evan Morikawa and
                  Alec Radford and
                  Matthew Knight and
                  Miles Brundage and
                  Mira Murati and
                  Katie Mayer and
                  Peter Welinder and
                  Bob McGrew and
                  Dario Amodei and
                  Sam McCandlish and
                  Ilya Sutskever and
                  Wojciech Zaremba},
  title        = {Evaluating Large Language Models Trained on Code},
  journal      = {CoRR},
  volume       = {abs/2107.03374},
  year         = {2021},
  eprinttype    = {arXiv},
  eprint       = {2107.03374},
  bibsource    = {dblp computer science bibliography, https://dblp.org}
}

@inproceedings{LiuXW023,
  author       = {Jiawei Liu and
                  Chunqiu Steven Xia and
                  Yuyao Wang and
                  Lingming Zhang},
  editor       = {Alice Oh and
                  Tristan Naumann and
                  Amir Globerson and
                  Kate Saenko and
                  Moritz Hardt and
                  Sergey Levine},
  title        = {Is Your Code Generated by ChatGPT Really Correct? Rigorous Evaluation
                  of Large Language Models for Code Generation},
  booktitle    = {Advances in Neural Information Processing Systems 36: Annual Conference
                  on Neural Information Processing Systems 2023, NeurIPS 2023, New Orleans,
                  LA, USA, December 10 - 16, 2023},
  year         = {2023},
  bibsource    = {dblp computer science bibliography, https://dblp.org}
}

@article{DongJJL24,
  author       = {Yihong Dong and
                  Xue Jiang and
                  Zhi Jin and
                  Ge Li},
  title        = {Self-Collaboration Code Generation via ChatGPT},
  journal      = {{ACM} Trans. Softw. Eng. Methodol.},
  volume       = {33},
  number       = {7},
  pages        = {189:1--189:38},
  year         = {2024},
  doi          = {10.1145/3672459},
  bibsource    = {dblp computer science bibliography, https://dblp.org}
}

@inproceedings{IslamAP24,
  author       = {Md. Ashraful Islam and
                  Mohammed Eunus Ali and
                  Md. Rizwan Parvez},
  editor       = {Lun{-}Wei Ku and
                  Andre Martins and
                  Vivek Srikumar},
  title        = {MapCoder: Multi-Agent Code Generation for Competitive Problem Solving},
  booktitle    = {Proceedings of the 62nd Annual Meeting of the Association for Computational
                  Linguistics (Volume 1: Long Papers), {ACL} 2024, Bangkok, Thailand,
                  August 11-16, 2024},
  pages        = {4912--4944},
  publisher    = {Association for Computational Linguistics},
  year         = {2024},
  doi          = {10.18653/V1/2024.ACL-LONG.269},
  bibsource    = {dblp computer science bibliography, https://dblp.org}
}

@inproceedings{IslamAP25,
  author       = {Md. Ashraful Islam and
                  Mohammed Eunus Ali and
                  Md. Rizwan Parvez},
  editor       = {Luis Chiruzzo and
                  Alan Ritter and
                  Lu Wang},
  title        = {CodeSim: Multi-Agent Code Generation and Problem Solving through Simulation-Driven
                  Planning and Debugging},
  booktitle    = {Findings of the Association for Computational Linguistics: {NAACL}
                  2025, Albuquerque, New Mexico, USA, April 29 - May 4, 2025},
  pages        = {5113--5139},
  publisher    = {Association for Computational Linguistics},
  year         = {2025},
  url          = {https://aclanthology.org/2025.findings-naacl.285/},
  bibsource    = {dblp computer science bibliography, https://dblp.org}
}

@inproceedings{ZhangLLSJ24,
  author       = {Kechi Zhang and
                  Jia Li and
                  Ge Li and
                  Xianjie Shi and
                  Zhi Jin},
  editor       = {Lun{-}Wei Ku and
                  Andre Martins and
                  Vivek Srikumar},
  title        = {CodeAgent: Enhancing Code Generation with Tool-Integrated Agent Systems
                  for Real-World Repo-level Coding Challenges},
  booktitle    = {Proceedings of the 62nd Annual Meeting of the Association for Computational
                  Linguistics (Volume 1: Long Papers), {ACL} 2024, Bangkok, Thailand,
                  August 11-16, 2024},
  pages        = {13643--13658},
  publisher    = {Association for Computational Linguistics},
  year         = {2024},
  url          = {https://doi.org/10.18653/v1/2024.acl-long.737},
  doi          = {10.18653/V1/2024.ACL-LONG.737},
  bibsource    = {dblp computer science bibliography, https://dblp.org}
}

@inproceedings{NiuL0022,
  author       = {Changan Niu and
                  Chuanyi Li and
                  Bin Luo and
                  Vincent Ng},
  editor       = {Luc De Raedt},
  title        = {Deep Learning Meets Software Engineering: {A} Survey on Pre-Trained
                  Models of Source Code},
  booktitle    = {Proceedings of the Thirty-First International Joint Conference on
                  Artificial Intelligence, {IJCAI} 2022, Vienna, Austria, 23-29 July
                  2022},
  pages        = {5546--5555},
  publisher    = {ijcai.org},
  year         = {2022},
  url          = {https://doi.org/10.24963/ijcai.2022/775},
  doi          = {10.24963/IJCAI.2022/775},
  bibsource    = {dblp computer science bibliography, https://dblp.org}
}

@inproceedings{FengGTDFGS0LJZ20,
  author       = {Zhangyin Feng and
                  Daya Guo and
                  Duyu Tang and
                  Nan Duan and
                  Xiaocheng Feng and
                  Ming Gong and
                  Linjun Shou and
                  Bing Qin and
                  Ting Liu and
                  Daxin Jiang and
                  Ming Zhou},
  editor       = {Trevor Cohn and
                  Yulan He and
                  Yang Liu},
  title        = {CodeBERT: {A} Pre-Trained Model for Programming and Natural Languages},
  booktitle    = {Findings of the Association for Computational Linguistics: {EMNLP}
                  2020, Online Event, 16-20 November 2020},
  series       = {Findings of {ACL}},
  volume       = {{EMNLP} 2020},
  pages        = {1536--1547},
  publisher    = {Association for Computational Linguistics},
  year         = {2020},
  url          = {https://doi.org/10.18653/v1/2020.findings-emnlp.139},
  doi          = {10.18653/V1/2020.FINDINGS-EMNLP.139},
  bibsource    = {dblp computer science bibliography, https://dblp.org}
}

@inproceedings{0034WJH21,
  author       = {Yue Wang and
                  Weishi Wang and
                  Shafiq R. Joty and
                  Steven C. H. Hoi},
  editor       = {Marie{-}Francine Moens and
                  Xuanjing Huang and
                  Lucia Specia and
                  Scott Wen{-}tau Yih},
  title        = {CodeT5: Identifier-aware Unified Pre-trained Encoder-Decoder Models
                  for Code Understanding and Generation},
  booktitle    = {Proceedings of the 2021 Conference on Empirical Methods in Natural
                  Language Processing, {EMNLP} 2021, Virtual Event / Punta Cana, Dominican
                  Republic, 7-11 November, 2021},
  pages        = {8696--8708},
  publisher    = {Association for Computational Linguistics},
  year         = {2021},
  url          = {https://doi.org/10.18653/v1/2021.emnlp-main.685},
  doi          = {10.18653/V1/2021.EMNLP-MAIN.685},
  bibsource    = {dblp computer science bibliography, https://dblp.org}
}

@inproceedings{RabinovichSK17,
  author       = {Maxim Rabinovich and
                  Mitchell Stern and
                  Dan Klein},
  editor       = {Regina Barzilay and
                  Min{-}Yen Kan},
  title        = {Abstract Syntax Networks for Code Generation and Semantic Parsing},
  booktitle    = {Proceedings of the 55th Annual Meeting of the Association for Computational
                  Linguistics, {ACL} 2017, Vancouver, Canada, July 30 - August 4, Volume
                  1: Long Papers},
  pages        = {1139--1149},
  publisher    = {Association for Computational Linguistics},
  year         = {2017},
  url          = {https://doi.org/10.18653/v1/P17-1105},
  doi          = {10.18653/V1/P17-1105},
  bibsource    = {dblp computer science bibliography, https://dblp.org}
}

@inproceedings{ParvezACRC21,
  author       = {Md. Rizwan Parvez and
                  Wasi Uddin Ahmad and
                  Saikat Chakraborty and
                  Baishakhi Ray and
                  Kai{-}Wei Chang},
  editor       = {Marie{-}Francine Moens and
                  Xuanjing Huang and
                  Lucia Specia and
                  Scott Wen{-}tau Yih},
  title        = {Retrieval Augmented Code Generation and Summarization},
  booktitle    = {Findings of the Association for Computational Linguistics: {EMNLP}
                  2021, Virtual Event / Punta Cana, Dominican Republic, 16-20 November,
                  2021},
  pages        = {2719--2734},
  publisher    = {Association for Computational Linguistics},
  year         = {2021},
  url          = {https://doi.org/10.18653/v1/2021.findings-emnlp.232},
  doi          = {10.18653/V1/2021.FINDINGS-EMNLP.232},
  bibsource    = {dblp computer science bibliography, https://dblp.org}
}

@article{LiAZMKMMALCLZZW23,
  author       = {Raymond Li and
                  Loubna Ben Allal and
                  Yangtian Zi and
                  Niklas Muennighoff and
                  Denis Kocetkov and
                  Chenghao Mou and
                  Marc Marone and
                  Christopher Akiki and
                  Jia Li and
                  Jenny Chim and
                  Qian Liu and
                  Evgenii Zheltonozhskii and
                  Terry Yue Zhuo and
                  Thomas Wang and
                  Olivier Dehaene and
                  Mishig Davaadorj and
                  Joel Lamy{-}Poirier and
                  Jo{\~{a}}o Monteiro and
                  Oleh Shliazhko and
                  Nicolas Gontier and
                  Nicholas Meade and
                  Armel Zebaze and
                  Ming{-}Ho Yee and
                  Logesh Kumar Umapathi and
                  Jian Zhu and
                  Benjamin Lipkin and
                  Muhtasham Oblokulov and
                  Zhiruo Wang and
                  Rudra Murthy V and
                  Jason T. Stillerman and
                  Siva Sankalp Patel and
                  Dmitry Abulkhanov and
                  Marco Zocca and
                  Manan Dey and
                  Zhihan Zhang and
                  Nour Fahmy and
                  Urvashi Bhattacharyya and
                  Wenhao Yu and
                  Swayam Singh and
                  Sasha Luccioni and
                  Paulo Villegas and
                  Maxim Kunakov and
                  Fedor Zhdanov and
                  Manuel Romero and
                  Tony Lee and
                  Nadav Timor and
                  Jennifer Ding and
                  Claire Schlesinger and
                  Hailey Schoelkopf and
                  Jan Ebert and
                  Tri Dao and
                  Mayank Mishra and
                  Alex Gu and
                  Jennifer Robinson and
                  Carolyn Jane Anderson and
                  Brendan Dolan{-}Gavitt and
                  Danish Contractor and
                  Siva Reddy and
                  Daniel Fried and
                  Dzmitry Bahdanau and
                  Yacine Jernite and
                  Carlos Mu{\~{n}}oz Ferrandis and
                  Sean Hughes and
                  Thomas Wolf and
                  Arjun Guha and
                  Leandro von Werra and
                  Harm de Vries},
  title        = {StarCoder: may the source be with you!},
  journal      = {Trans. Mach. Learn. Res.},
  volume       = {2023},
  year         = {2023},
  url          = {https://openreview.net/forum?id=KoFOg41haE},
  bibsource    = {dblp computer science bibliography, https://dblp.org}
}

@article{abs-2308-12950,
  author       = {Baptiste Rozi{\`{e}}re and
                  Jonas Gehring and
                  Fabian Gloeckle and
                  Sten Sootla and
                  Itai Gat and
                  Xiaoqing Ellen Tan and
                  Yossi Adi and
                  Jingyu Liu and
                  Tal Remez and
                  J{\'{e}}r{\'{e}}my Rapin and
                  Artyom Kozhevnikov and
                  Ivan Evtimov and
                  Joanna Bitton and
                  Manish Bhatt and
                  Cristian Canton{-}Ferrer and
                  Aaron Grattafiori and
                  Wenhan Xiong and
                  Alexandre D{\'{e}}fossez and
                  Jade Copet and
                  Faisal Azhar and
                  Hugo Touvron and
                  Louis Martin and
                  Nicolas Usunier and
                  Thomas Scialom and
                  Gabriel Synnaeve},
  title        = {Code Llama: Open Foundation Models for Code},
  journal      = {CoRR},
  volume       = {abs/2308.12950},
  year         = {2023},
  url          = {https://doi.org/10.48550/arXiv.2308.12950},
  doi          = {10.48550/ARXIV.2308.12950},
  eprinttype    = {arXiv},
  eprint       = {2308.12950},
  bibsource    = {dblp computer science bibliography, https://dblp.org}
}

@article{abs-2401-14196,
  author       = {Daya Guo and
                  Qihao Zhu and
                  Dejian Yang and
                  Zhenda Xie and
                  Kai Dong and
                  Wentao Zhang and
                  Guanting Chen and
                  Xiao Bi and
                  Y. Wu and
                  Y. K. Li and
                  Fuli Luo and
                  Yingfei Xiong and
                  Wenfeng Liang},
  title        = {DeepSeek-Coder: When the Large Language Model Meets Programming -
                  The Rise of Code Intelligence},
  journal      = {CoRR},
  volume       = {abs/2401.14196},
  year         = {2024},
  url          = {https://doi.org/10.48550/arXiv.2401.14196},
  doi          = {10.48550/ARXIV.2401.14196},
  eprinttype    = {arXiv},
  eprint       = {2401.14196},
  bibsource    = {dblp computer science bibliography, https://dblp.org}
}

@article{abs-2409-12186,
  author       = {Binyuan Hui and
                  Jian Yang and
                  Zeyu Cui and
                  Jiaxi Yang and
                  Dayiheng Liu and
                  Lei Zhang and
                  Tianyu Liu and
                  Jiajun Zhang and
                  Bowen Yu and
                  Kai Dang and
                  An Yang and
                  Rui Men and
                  Fei Huang and
                  Xingzhang Ren and
                  Xuancheng Ren and
                  Jingren Zhou and
                  Junyang Lin},
  title        = {Qwen2.5-Coder Technical Report},
  journal      = {CoRR},
  volume       = {abs/2409.12186},
  year         = {2024},
  url          = {https://doi.org/10.48550/arXiv.2409.12186},
  doi          = {10.48550/ARXIV.2409.12186},
  eprinttype    = {arXiv},
  eprint       = {2409.12186},
  bibsource    = {dblp computer science bibliography, https://dblp.org}
}

@article{DongDJLLJ25,
  author       = {Yihong Dong and
                  Jiazheng Ding and
                  Xue Jiang and
                  Ge Li and
                  Zhuo Li and
                  Zhi Jin},
  title        = {CodeScore: Evaluating Code Generation by Learning Code Execution},
  journal      = {{ACM} Trans. Softw. Eng. Methodol.},
  volume       = {34},
  number       = {3},
  pages        = {77:1--77:22},
  year         = {2025},
  url          = {https://doi.org/10.1145/3695991},
  doi          = {10.1145/3695991},
  bibsource    = {dblp computer science bibliography, https://dblp.org}
}

@article{abs-2108-07732,
  author       = {Jacob Austin and
                  Augustus Odena and
                  Maxwell I. Nye and
                  Maarten Bosma and
                  Henryk Michalewski and
                  David Dohan and
                  Ellen Jiang and
                  Carrie J. Cai and
                  Michael Terry and
                  Quoc V. Le and
                  Charles Sutton},
  title        = {Program Synthesis with Large Language Models},
  journal      = {CoRR},
  volume       = {abs/2108.07732},
  year         = {2021},
  url          = {https://arxiv.org/abs/2108.07732},
  eprinttype    = {arXiv},
  eprint       = {2108.07732},
  bibsource    = {dblp computer science bibliography, https://dblp.org}
}

@inproceedings{JainHGLYZWSSS25,
  author       = {Naman Jain and
                  King Han and
                  Alex Gu and
                  Wen{-}Ding Li and
                  Fanjia Yan and
                  Tianjun Zhang and
                  Sida Wang and
                  Armando Solar{-}Lezama and
                  Koushik Sen and
                  Ion Stoica},
  title        = {LiveCodeBench: Holistic and Contamination Free Evaluation of Large
                  Language Models for Code},
  booktitle    = {The Thirteenth International Conference on Learning Representations,
                  {ICLR} 2025, Singapore, April 24-28, 2025},
  publisher    = {OpenReview.net},
  year         = {2025},
  url          = {https://openreview.net/forum?id=chfJJYC3iL},
  bibsource    = {dblp computer science bibliography, https://dblp.org}
}

@article{YangZCZZC24,
  author       = {Guang Yang and
                  Yu Zhou and
                  Xiang Chen and
                  Xiangyu Zhang and
                  Terry Yue Zhuo and
                  Taolue Chen},
  title        = {Chain-of-Thought in Neural Code Generation: From and for Lightweight
                  Language Models},
  journal      = {{IEEE} Trans. Software Eng.},
  volume       = {50},
  number       = {9},
  pages        = {2437--2457},
  year         = {2024},
  url          = {https://doi.org/10.1109/TSE.2024.3440503},
  doi          = {10.1109/TSE.2024.3440503},
  bibsource    = {dblp computer science bibliography, https://dblp.org}
}

@article{JiangDWFSLJJ24,
  author       = {Xue Jiang and
                  Yihong Dong and
                  Lecheng Wang and
                  Zheng Fang and
                  Qiwei Shang and
                  Ge Li and
                  Zhi Jin and
                  Wenpin Jiao},
  title        = {Self-Planning Code Generation with Large Language Models},
  journal      = {{ACM} Trans. Softw. Eng. Methodol.},
  volume       = {33},
  number       = {7},
  pages        = {182:1--182:30},
  year         = {2024},
  url          = {https://doi.org/10.1145/3672456},
  doi          = {10.1145/3672456},
  bibsource    = {dblp computer science bibliography, https://dblp.org}
}

@inproceedings{ZhangLSCSY25,
  author       = {Xiaoqing Zhang and
                  Yuhan Liu and
                  Flood Sung and
                  Xiuying Chen and
                  Shuo Shang and
                  Rui Yan},
  editor       = {Wanxiang Che and
                  Joyce Nabende and
                  Ekaterina Shutova and
                  Mohammad Taher Pilehvar},
  title        = {Thinking Before Running! Efficient Code Generation with Thorough Exploration
                  and Optimal Refinement},
  booktitle    = {Findings of the Association for Computational Linguistics, {ACL} 2025,
                  Vienna, Austria, July 27 - August 1, 2025},
  pages        = {23268--23281},
  publisher    = {Association for Computational Linguistics},
  year         = {2025},
  url          = {https://aclanthology.org/2025.findings-acl.1195/},
  bibsource    = {dblp computer science bibliography, https://dblp.org}
}

@inproceedings{khan-etal-2024-xcodeeval,
    title = "{XC}ode{E}val: An Execution-based Large Scale Multilingual Multitask Benchmark for Code Understanding, Generation, Translation and Retrieval",
    author = "Khan, Mohammad Abdullah Matin  and
      Bari, M Saiful  and
      Do, Xuan Long  and
      Wang, Weishi  and
      Parvez, Md Rizwan  and
      Joty, Shafiq",
    editor = "Ku, Lun-Wei  and
      Martins, Andre  and
      Srikumar, Vivek",
    booktitle = "Proceedings of the 62nd Annual Meeting of the Association for Computational Linguistics (Volume 1: Long Papers)",
    month = aug,
    year = "2024",
    address = "Bangkok, Thailand",
    publisher = "Association for Computational Linguistics",
    url = "https://aclanthology.org/2024.acl-long.367/",
    doi = "10.18653/v1/2024.acl-long.367",
    pages = "6766--6805"
}

@inproceedings{paircoder,
  author       = {Huan Zhang and
                  Wei Cheng and
                  Yuhan Wu and
                  Wei Hu},
  editor       = {Vladimir Filkov and
                  Baishakhi Ray and
                  Minghui Zhou},
  title        = {A Pair Programming Framework for Code Generation via Multi-Plan Exploration
                  and Feedback-Driven Refinement},
  booktitle    = {Proceedings of the 39th {IEEE/ACM} International Conference on Automated
                  Software Engineering, {ASE} 2024, Sacramento, CA, USA, October 27
                  - November 1, 2024},
  pages        = {1319--1331},
  publisher    = {{ACM}},
  year         = {2024},
  url          = {https://doi.org/10.1145/3691620.3695506},
  doi          = {10.1145/3691620.3695506},
  bibsource    = {dblp computer science bibliography, https://dblp.org}
}

@article{brown2024large,
  title   = {Large Language Monkeys: Scaling Inference Compute with Repeated Sampling},
  author  = {Brown, Bradley and Juravsky, Jordan and Ehrlich, Ryan and Clark, Ronald and Le, Quoc V. and R{\'e}, Christopher and Mirhoseini, Azalia},
  journal = {arXiv preprint arXiv:2407.21787},
  year    = {2024},
  doi     = {10.48550/arXiv.2407.21787},
  url     = {https://arxiv.org/abs/2407.21787}
}

@inproceedings{shinn2023reflexion,
  author       = {Noah Shinn and
                  Federico Cassano and
                  Ashwin Gopinath and
                  Karthik Narasimhan and
                  Shunyu Yao},
  editor       = {Alice Oh and
                  Tristan Naumann and
                  Amir Globerson and
                  Kate Saenko and
                  Moritz Hardt and
                  Sergey Levine},
  title        = {Reflexion: language agents with verbal reinforcement learning},
  booktitle    = {Advances in Neural Information Processing Systems 36: Annual Conference
                  on Neural Information Processing Systems 2023, NeurIPS 2023, New Orleans,
                  LA, USA, December 10 - 16, 2023},
  year         = {2023},
  url          = {http://papers.nips.cc/paper\_files/paper/2023/hash/1b44b878bb782e6954cd888628510e90-Abstract-Conference.html},
  bibsource    = {dblp computer science bibliography, https://dblp.org}
}
\appendix

\section{Error Analysis with Examples}
To better understand the limitations of CollabCoder, we analyze cases where the framework fails or produces partially correct solutions. Our error analysis highlights recurring failure modes, including missing functionality, incomplete edge-case handling, semantic misalignment with the specification, and performance issues:

\subsection{Case Study 1: Binary Search Tree (from framework example)} 

As illustrated in Figure \ref{fig2}, CollabCoder initially generated a Binary Search Tree class with insert and search methods but omitted the delete method and null handling. This led to 29 out of 47 test failures in the first iteration. The collaborative debug agent localized errors, prompting plan refinement to add delete functionality and edge-case handling. After two iterations, the solution improved significantly (83\% pass rate) but still missed advanced balancing and performance optimization. This example illustrates how CollabCoder effectively reduces reliance on superficial trial-and-error by progressively incorporating missing components. However, conceptual leaps (e.g., balanced tree optimization) remain difficult.

\subsection{Case Study 2: MBPP Task 612 - Merging Sublists} 
``Write a function to merge the first and last elements separately in a list of lists''

\textbf{Expected Behavior:} The function should also generalize to cases where sublists contain more than two elements. For example:
\[
\begin{aligned}
\big[ &["x","y","z"],\; ["a","b","c"],\; ["m","n","o"] \big] \\
&\longrightarrow \\
\big[ &["x","a","m"],\; ["y","b","n"],\; ["z","c","o"] \big]
\end{aligned}
\]

\textbf{Generated Output:}
\begin{lstlisting}[language=Python]
def merge(lst: List[List[Any]]) -> List[List[Any]]:
    if not lst:
        return [[], []]
    first_elements = []
    second_elements = []
    for sublist in lst:
        if len(sublist) != 2:
            raise ValueError("Each sublist must contain exactly two elements.")
        first_elements.append(sublist[0])
        second_elements.append(sublist[1])
    return [first_elements, second_elements]
\end{lstlisting}

\textbf{Failure Analysis:}  
The code worked for sublists of exactly two elements but failed on test cases where sublists contained three elements. Specifically:
\begin{itemize}
    \item \text{CollabCoder} enforced a restrictive assumption (\texttt{len(sublist) != 2} $\rightarrow$ error), contradicting the broader specification.
    \item Subsequent debugging iterations failed to revise the plan for variable-length sublists.
\end{itemize}

\textbf{Insight:} This case reveals a key limitation: CollabCoder can overfit to an overly strict interpretation of the specification during plan formation. Moreover, its debugging mechanism struggles to generalize when the problem requires extending functionality beyond initial assumptions.

In general, while CollabCoder substantially improves over baselines in robustness and adaptability, systematic error analysis highlights the need for stronger semantic alignment between problem specification and evolving plans, particularly in tasks that require flexible generalization.



\section{Experimental Details}
\subsection{Baselines Description}
\label{app:baseline}
Direct Prompting requires language models to generate code directly, without any additional instructions or intermediate reasoning steps, leveraging their intrinsic problem-solving abilities. CoT~\cite{YangZCZZC24} encourages models to generate intermediate natural language reasoning to guide code generation, while Self-Planning~\cite{JiangDWFSLJJ24} separates the process into a planning phase and an implementation phase. MapCoder~\cite{IslamAP24} adopts a multi-agent architecture covering example retrieval, planning, coding, and debugging, and CodeSIM~\cite{IslamAP25} extends this framework by incorporating simulated input/output execution for iterative verification. ThinkCoder~\cite{ZhangLSCSY25} employs a streamlined two-agent design with an Exploration Agent for generating diverse solutions and a CodeVerifier Agent for independent evaluation using a test pool.
\subsection{Dataset Description}

For \textbf{HumanEval} and \textbf{MBPP}, we follow the standardized preprocessing protocol adopted in prior work~\cite{IslamAP24} to ensure fair and reproducible evaluation. After preprocessing, the resulting benchmark consists of 164 HumanEval problems and 397 MBPP problems. Their extended counterparts, HumanEval-ET (HE-ET) and MBPP-ET, augment each original problem with additional test cases, enabling a more stringent assessment of functional correctness and robustness under out-of-distribution inputs~\cite{DongDJLLJ25}. To evaluate performance on more challenging, competition-style programming tasks, we further employ \textbf{LiveCodeBench} and \textbf{xCodeEval}, which are designed to reflect real-world coding difficulty and temporal data distribution. In particular, we use the most recent release of LiveCodeBench (version~6), which contains 175 code generation problems collected between May 2023 and April 2025\footnote{\url{https://huggingface.co/datasets/livecodebench/code_generation_lite}}. Following common practice for consistent evaluation, we exclude problems formulated in a functional input--output style and retain only 112 Stdin/Stdout-based tasks, which align with the execution-based evaluation protocol used in this work.

Similar to prior methods, we use publicly available test cases across all benchmarks only for implementation sanity checking and debugging during development. The reported results are obtained by executing the final generated code exclusively on the official hidden/private test cases provided by the corresponding benchmarks.

To enable execution-based evaluation on these datasets, the LLM is required to generate code that strictly follows a predefined coding template compatible with the corresponding evaluation framework. This requirement ensures that the generated code can be executed automatically and evaluated consistently across different benchmarks.

HumanEval and MBPP adopt a functional coding paradigm, where the model is expected to generate a single Python function according to the function signature and problem description provided in the prompt, and the generated function is then directly invoked by the evaluator.

\textbf{Example coding template for HumanEval}:
\begin{Verbatim}[fontsize=\footnotesize,breaklines=true,breakanywhere=true,frame=single,framesep=2mm]
def truncate_number(number: float) -> float:
    """ Given a positive floating point number, it can be decomposed into
    and integer part (largest integer smaller than given number) and decimals
    (leftover part always smaller than 1).

    Return the decimal part of the number.
    >>> truncate_number(3.5)
    0.5
    """
\end{Verbatim}

\textbf{Example coding template for MBPP}:
\begin{Verbatim}[fontsize=\footnotesize,breaklines=true,breakanywhere=true,frame=single,framesep=2mm]
from typing import List

def heap_queue_largest(nums: List[int], n: int) -> List[int]:
    """
    Write a function to find the n largest integers from a given list of numbers,
    returned in descending order.
    """
\end{Verbatim}

In contrast, xCodeEval and LiveCodeBench follow a stdin/stdout-based coding format. In this setting, the model is required to generate a complete program that reads inputs from standard input, processes them according to the problem specification, and writes the correct outputs to standard output. The entire program is executed as a script during evaluation, which more closely reflects real-world programming contest environments. 

\textbf{Example problem from xCodeEval}:
\begin{Verbatim}[
  fontsize=\footnotesize,
  breaklines=true,
  breakanywhere=true,
  breaksymbolleft={},
  breaksymbolright={},
  frame=single,
  framesep=2mm
]
Title: Make All Numbers Odd

Problem Description:
There are n positive integers a_1, a_2, ..., a_n. In one move, you can choose
an even value c and divide by two all elements equal to c. The goal is to find
the minimum number of moves required to make all numbers odd.
Input:
The first line contains an integer t, the number of test cases.
For each test case, the first line contains an integer n.
The second line contains n positive integers a_1, a_2, ..., a_n.
Output:
For each test case, output the minimum number of moves required.
Constraints:
1 <= t <= 10^4,
1 <= n <= 2*10^5 (sum of n over all test cases),
1 <= a_i <= 10^9.
Sample Input:
4
6
40 6 40 3 20 1
1
1024
4
2 4 8 16
3
3 1 7
Sample Output:
4
10
4
0
\end{Verbatim}

\textbf{Example problem from LiveCodeBench}:
\begin{Verbatim}[
  fontsize=\footnotesize,
  breaklines=true,
  breakanywhere=true,
  breaksymbolleft={},
  breaksymbolright={},
  frame=single,
  framesep=2mm
]
Title: ABC400 Ceremony

Problem Description:
In the ceremony commemorating ABC400, we want to arrange 400 people in a
rectangular formation with A rows and B columns. Given A, determine a positive integer B such that AB = 400. If no such B exists, output -1.

Input:
A single integer A.

Output:
Output a single integer B, or -1 if no valid arrangement exists.

Constraints:
1 <= A <= 400.

Sample Input:
10

Sample Output:
40
\end{Verbatim}

\subsection{Metrics}
\label{app:metric}
\paragraph{Update Rate.} Plan Update Rate (PU Rate) and Code Update Rate (CU Rate) measure the proportion of plan-level and code-level update decisions, respectively, made by the CDM module over the total number of executed iterations. Unlike raw update counts, which are often biased by the number of samples in a dataset, update rates provide a normalized measure that enables fair comparison across datasets of different sizes. Importantly, these metrics also account for iterations in which the CDM module decides not to perform any update, thereby reflecting the full decision space of the system rather than only corrective actions. The metrics on a dataset 
$\mathcal{S}$ are computed as follows:
\begin{equation}
\begin{aligned}
    \text{\textbf{PU  Rate}} = \frac{\text{\textbf{PU}}}{\text{\textbf{PU}} + \text{\textbf{CU}} + |\mathcal{S}|}, 
    \\
     \text{\textbf{CU Rate}} = \frac{\text{\textbf{CU}}}{\text{\textbf{PU}} + \text{\textbf{CU}} + |\mathcal{S}|}.
\end{aligned}
\end{equation}    
Here, \textbf{PU} and \textbf{CU} denote the total numbers of plan-level and code-level updates, respectively, aggregated over all samples in the dataset. $|\mathcal{S}|$ represents the size of the dataset $\mathcal{S}$. The denominator thus corresponds to the total number of executed iterations across all samples in $\mathcal{S}$.
\begin{figure*}[h]
    \centering
    \includegraphics[width=\linewidth]{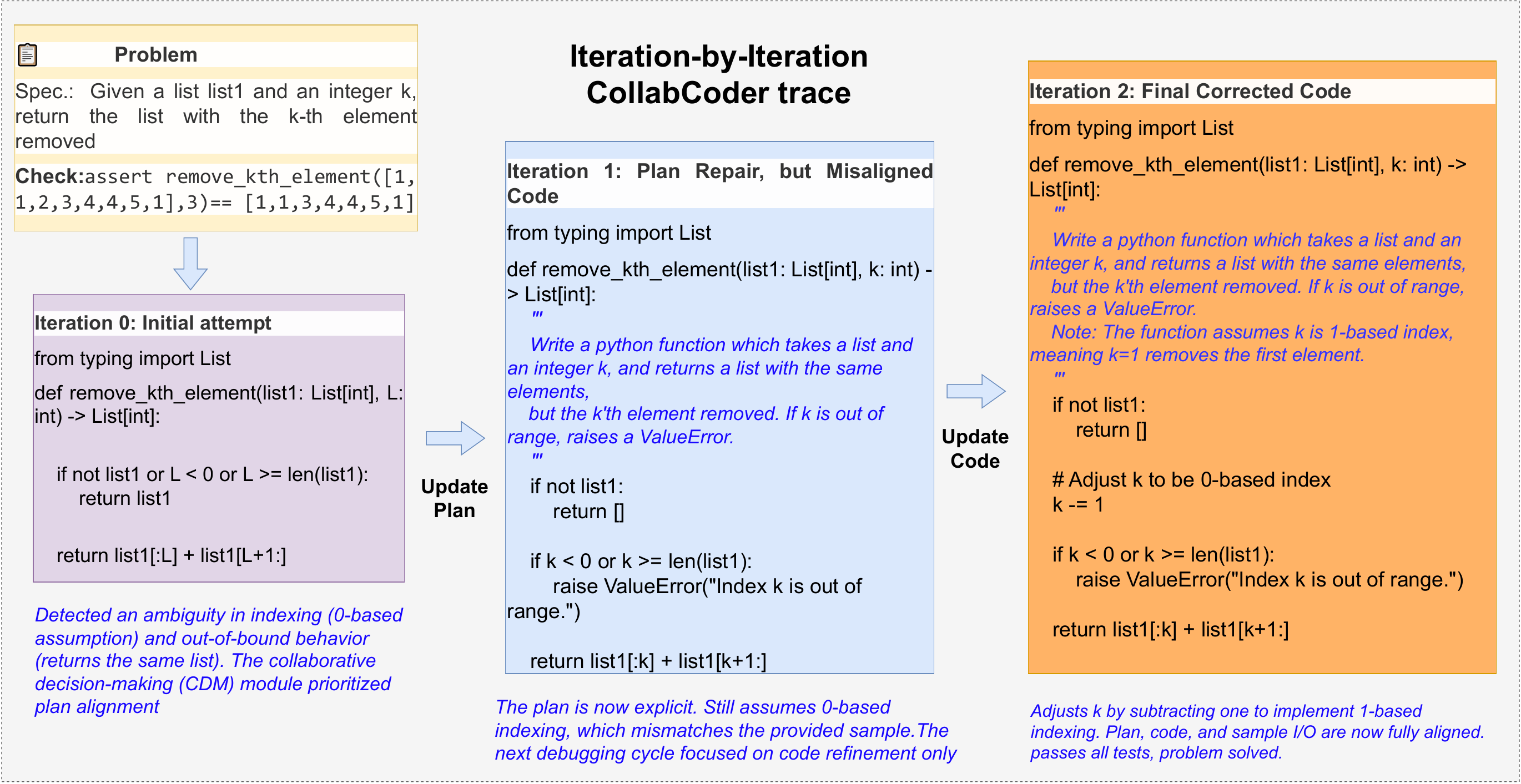}
    \caption{An example of self-improving debugging.}
    \label{fig:self_example}
\end{figure*}
\paragraph{Token Input/Output.} 

Token Input/Output (Token I/O) is defined as the combination of two complementary metrics: Token Input and Token Output. These metrics are computed as averages over the entire dataset and represent the average number of input and output tokens required to execute a single sample. The general formulation for computing Token Input and Token Output on a dataset $\mathcal{S}$ is as follows:
\begin{equation}
    \text{\textbf{Token Avg}} =     \frac{\sum_{i \in  \mathcal{S}}\sum_{k=1}^{a_i}t_k}{|\mathcal{S}|}.
\end{equation}
Here, \(a_i\) denotes the total number of API calls made during the execution of sample \(i\), and \(t_k\) denotes the number of tokens (either input or output) consumed in the \(k\)-th API call. Overall, Token I/O serves as an objective measure of framework efficiency by capturing the average token consumption required to solve a sample.

\section{Additional Analysis and Experiments}
\subsection{Results on Frontier Backbone LLMs}

\begin{table}[ht]
\centering
\small
\resizebox{\columnwidth}{!}{
\begin{tabular}{llcccc}
\toprule
\textbf{Dataset} & \textbf{Method} & \textbf{Pass@1} & \textbf{Token In} & \textbf{Token Out} & \textbf{API} \\
\midrule

\rowcolor{headergray}
\multicolumn{6}{l}{\textbf{Backbone LLM: GPT-5.2}} \\

\multirow{3}{*}{MBPP}
& MapCoder & 93.94 & 2717.31 & \phantom{0}3346.24 & \phantom{0}5.04 \\
& CodeSIM & 94.74 & 1998.49 & \phantom{0}2861.05 & \phantom{0}4.60 \\
& \textbf{CollabCoder (Ours)} & 95.21 & 2293.08 & \phantom{0}2377.44 & \phantom{0}3.52 \\

\midrule

\multirow{3}{*}{LCB}
& MapCoder & 63.39 & 19691.38 & 12469.18 & 14.82 \\
& CodeSIM & 64.29 & 18294.05 & \phantom{0}7555.67 & 13.13 \\
& \textbf{CollabCoder (Ours)} & 65.18 & 12047.65 & \phantom{0}4293.60 & \phantom{0}9.66 \\

\midrule

\rowcolor{headergray}
\multicolumn{6}{l}{\textbf{Backbone LLM: Qwen3-Coder-Next}} \\

\multirow{3}{*}{MBPP}
& MapCoder & 89.42 &	5733.25 &	\phantom{0}2913.57 &	10.37 \\
& CodeSIM & 91.18 &	2329.11 &	\phantom{0}2679.33 &	\phantom{0}5.32 \\
& \textbf{CollabCoder (Ours)} & 92.69 &	2551.48 &	\phantom{0}1437.78	& \phantom{0}4.72 \\

\midrule

\multirow{3}{*}{LCB}
& MapCoder & 50.89	& 24368.88 &	14393.9 &	16.26 \\
& CodeSIM & 52.68 &	18811.83 &	12738.6 &	14.20 \\
& \textbf{CollabCoder (Ours)} & 55.38 &	14536.74 &	\phantom{0}4121.2 &	10.74 \\

\bottomrule
\end{tabular}
}
\caption{Performance and Efficiency on Frontier LLMs}
\label{tab:extend_main}
\end{table}

CollabCoder’s advantages are not confined to strengthening small- and medium-scale backbones. To further evaluate its generality under stronger LLM backbones, we extend Table~\ref{tab:main} by incorporating two frontier models, GPT-5.2 and Qwen3-Coder-Next (80B), and report results on both the MBPP and LCB benchmarks. As shown in Table~\ref{tab:extend_main}, the overall trend remains consistent with that observed in the main experiments: CollabCoder achieves the best Pass@1 across both benchmarks and both backbones, while also maintaining clear efficiency advantages. Although the accuracy gaps become smaller than those observed with small- and medium-scale backbones, this is expected due to the natural ceiling effect as backbone capability improves. Even under these stronger settings, CollabCoder still attains the highest Pass@1 for both GPT-5.2 and Qwen3-Coder-Next, while reducing API calls and output token usage relative to prior multi-agent baselines. This pattern is particularly notable for frontier backbones, where dynamic plan--code co-evolution continues to provide a favorable balance between effectiveness and efficiency.

Overall, these results suggest that the benefits of CollabCoder extend beyond mid-scale models. While the absolute accuracy margin naturally narrows as backbone capability increases, the framework remains consistently competitive in accuracy and particularly strong in inference efficiency, indicating good scalability to frontier LLM backbones.

\subsection{Self-improving Debugging Analysis}
\label{debug_analysis}
\begin{table}[ht]
\centering
\small
\resizebox{\columnwidth}{!}{
\begin{tabular}{llc}
\toprule
\textbf{Method} & \textbf{Budget Setting} & \textbf{Solved (out of 90)} \\
\midrule
\multirow{5}{*}{Best-of-N}
& $N=5$  & 32 \\
& $N=10$ & 33 \\
& $N=15$ & 33 \\
& $N=20$ & 33 \\
& $N=25$ & 33 \\
\midrule
\multirow{4}{*}{Reflexion}
& $t=5$  & 35 \\
& $t=10$ & 37 \\
& $t=15$ & 37 \\
& $t=20$ & 37 \\
\midrule
\multirow{5}{*}{CollabCoder}
& $t=1$  & 32 \\
& $t=2$  & 35 \\
& $t=3$  & 37 \\
& $t=4$  & 38 \\
& $t=5$  & 44 \\
\bottomrule
\end{tabular}}
\caption{Comparison with limited inference-time baselines on the first 90 problems of LiveCodeBench using GPT-4o-mini.}
\label{tab:limited_baseline}
\end{table}

Figure~\ref{fig:self_example} provides an intuitive example of CollabCoder's self-improving debugging process. Starting from an initial solution with a semantic mismatch (i.e., an incorrect 0-based indexing assumption), the CDM module first identifies that the error originates from plan-level misalignment rather than a purely local code bug. It therefore repairs the plan to make the specification explicit in the first iteration, and then refines the code in the second iteration by adjusting the implementation to 1-based indexing. The example illustrates the core design of CollabCoder: iterative refinement is not merely repeated debugging, but a structured process that explicitly distinguishes between plan-level and implementation-level errors.

Furthermore, we investigate a complementary question: should inference-time self-improvement be allocated primarily to breadth or to depth? On the breadth side, a representative strategy is Best-of-$N$ sampling, which increases test-time compute by generating multiple candidate programs and selecting the one that passes the largest number of sample test cases for final submission \cite{brown2024large}. On the depth side, a representative line of work is trial-and-error debugging with feedback, exemplified by Reflexion \cite{shinn2023reflexion}, which iteratively improves subsequent attempts through verbal feedback. CollabCoder belongs to the latter category, but it is further enhanced with reasoning trajectory accumulation and collaborative decision making, enabling more effective debugging while reducing repeated mistakes across iterations.

To study this question, we compare CollabCoder with two simpler inference-time baselines, Best-of-$N$ and Reflexion, on the first 90 problems of LiveCodeBench using GPT-4o-mini. The results in Table~\ref{tab:limited_baseline} show clear differences in how performance evolves with additional budget. Best-of-$N$ saturates very early, improving only from 32 to 33 solved problems when increasing $N$ from 5 to 25. Reflexion yields stronger initial gains, but also plateaus quickly at 37/90 even as the number of reflection rounds increases. In contrast, CollabCoder improves steadily across iterations, from 32/90 at $t{=}1$ to 44/90 at $t{=}5$, ultimately surpassing both baselines by a clear margin.

These results suggest that CollabCoder's gains are not simply due to using more inference budget. Rather, they come from structured plan-code co-evolution, which allows the framework to revise flawed high-level strategies instead of repeatedly refining code under an incorrect plan. This distinction is especially important on complex programming problems, where persistent failures often arise from strategy errors that simpler self-improvement schemes cannot explicitly correct.

\subsection{Additional Self-Improving Analysis on xCodeEval}
\label{sec:additional_self_improve_xcodeeval}
\begin{figure}[ht]
    \centering
    \includegraphics[width=0.98\linewidth]{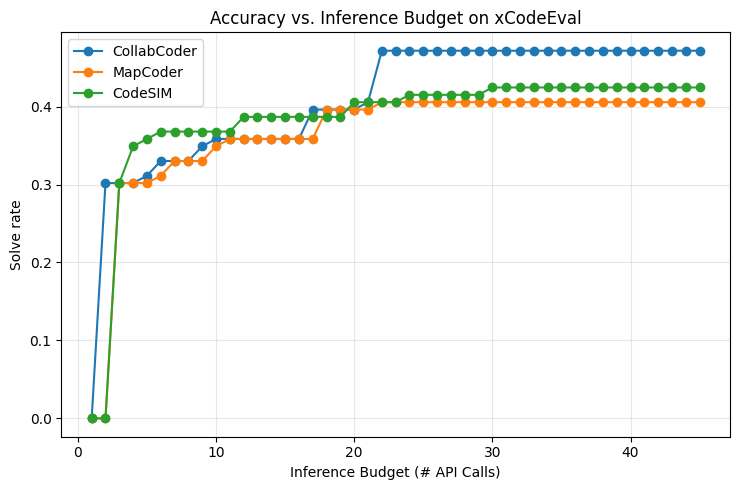}
    \caption{Accuracy vs.\ inference budget on xCodeEval benchmark.}
    \label{fig:accuracy_budget_xcodeeval}
\end{figure}

Figure~\ref{fig:accuracy_budget_xcodeeval} presents an additional self-improving analysis of different debugging strategies on the xCodeEval benchmark, focusing on how solution accuracy evolves with increasing inference budgets. 
At a low inference budget of 10 API calls, CollabCoder attains a solve rate of 35.85\%, which is comparable to MapCoder (34.91\%) and CodeSIM (36.79\%), indicating a relatively small early advantage on this benchmark. However, as the inference budget increases, CollabCoder exhibits a substantially steeper improvement trajectory. With approximately 20-25 API calls, CollabCoder rapidly converges to a higher solve rate of 47.16\%, a performance level that MapCoder and CodeSIM do not reach even with significantly larger inference budgets.

\subsection{CDM Analysis}
\label{CDM_analysis}
In this section, we present detailed statistics on plan-level and code-level updates observed during the execution of our pipeline across three benchmarks, namely HumanEval, MBPP, and LiveCodeBench, as shown in Figure~\ref{fig:cdm_analysis_rate}. The corresponding numerical results are reported in Table~\ref{tab:cdm_analysis}.
\begin{table}[ht]
\centering
\resizebox{\columnwidth}{!}{%
\begin{tabular}{lccccc}
\toprule
\multirow{2}{*}{\textbf{Model}} 
& \multirow{2}{*}{\textbf{Dataset}} 
& \multicolumn{2}{c}{\textbf{Plan Update}} 
& \multicolumn{2}{c}{\textbf{Code Update}} \\
\cmidrule(lr){3-4} \cmidrule(lr){5-6}
& & 
Quantity & Rate (\%) &
Quantity & Rate (\%) \\
\midrule
\multirow{3}{*}{Seed-Coder-8B} 
& HumanEval & 12 & 04.72 & 78 & 30.71 \\
\cmidrule(lr){2-2}
& MBPP      & 55 & 07.90 & 244 & 35.06 \\
\cmidrule(lr){2-2}
& LiveCodeBench  & 128 & 22.98 & 317 & 56.91 \\
\midrule
\multirow{3}{*}{Qwen2.5-Coder-32B} 
& HumanEval & 14 & 06.76 & 29 & 14.01 \\
\cmidrule(lr){2-2}
& MBPP & 64 & 11.74 & 84 & 15.41 \\
\cmidrule(lr){2-2}
& LiveCodeBench  & 80 & 22.16 & 169 & 46.81 \\
\midrule
\multirow{3}{*}{GPT-4o mini} 
& HumanEval & 70 & 27.78 & 18 & 07.14 \\
\cmidrule(lr){2-2}
& MBPP      & 167 & 27.11 & 52 & 08.44 \\
\cmidrule(lr){2-2}
& LiveCodeBench  & 231 & 57.46 & 59 & 14.68 \\
\bottomrule
\end{tabular}
}
\caption{A detailed report of the quantity and update rate of plan and code updates for different backbone models on HumanEval, MBPP, and LiveCodeBench.}
\label{tab:cdm_analysis}
\end{table}

\subsection{Hyperparameter Selection and Sensitivity of Trust Weights}

\begin{figure}[h]
    \centering
    \includegraphics[width=\linewidth]{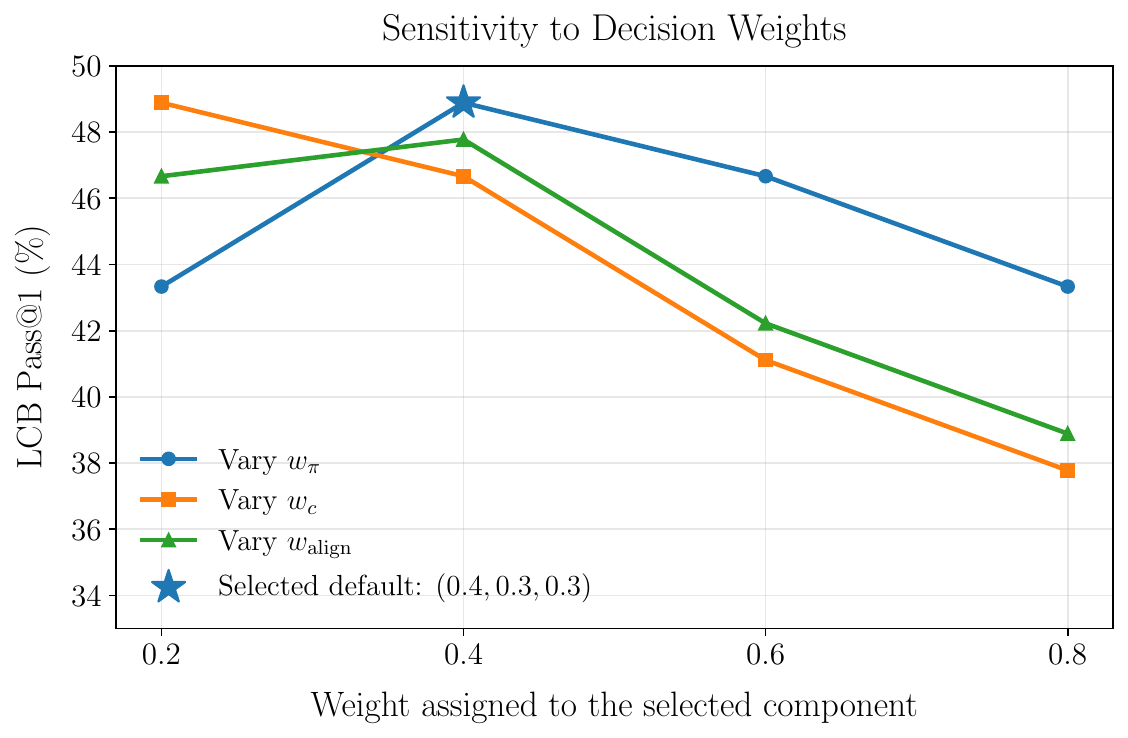}
    \caption{Sensitivity of CollabCoder to Trust Weights.
}
    \label{fig:sensitivity_weights}
\end{figure}

In this section, we analyze the sensitivity of the weighting coefficients $w_\pi$, $w_c$, and $w_{\text{align}}$ in Equation~\ref{eq:cdm}, which control the relative contributions of the plan-level, code-level, and plan-code alignment analyses in the CDM module. To conduct this analysis, for each component $x \in \{\pi, c, \text{align}\}$, we vary $w_x \in \{0.2, 0.4, 0.6, 0.8\}$ and set the remaining two weights as $w_y = w_z = (1 - w_x)/2$, ensuring that the three weights always sum to 1.

To examine robustness, we evaluate the resulting 12 normalized weight configurations on a 90-problem subset of LiveCodeBench using GPT-4o-mini as the backbone. As shown in Figure~\ref{fig:sensitivity_weights}, CollabCoder remains stable under moderate perturbations of the trust weights: several balanced configurations achieve comparable performance, while performance degrades mainly when a single signal becomes overly dominant. The default configuration used in the main experiments, $(w_\pi, w_c, w_{\text{align}}) = (0.4, 0.3, 0.3)$, introduces a mild bias toward plan-level analysis, encouraging correction of flawed high-level strategies while still preserving sufficient capacity for code-level and alignment-based refinement. Empirically, this setting achieves 48.9\% Pass@1 (44/90), tying for the best performance with $(0.4, 0.2, 0.4)$.

Notably, code-heavy configurations lead to the most pronounced drop in performance. For example, $(0.1, 0.8, 0.1)$ achieves only 37.8\% Pass@1, suggesting that excessive reliance on code-level refinement may suppress necessary revisions at the planning level. Alignment-heavy settings also reduce performance, whereas a moderate emphasis on plan-level analysis remains consistently effective across configurations.

Overall, these results indicate that CollabCoder does not rely on delicate weight tuning. Instead, it remains robust across a reasonably broad range of trust-weight choices, while benefiting most from configurations that maintain a balanced yet slightly plan-oriented emphasis.

\section{Implementation Details}
\subsection{Detailed Algorithm of CollabCoder}
\label{app:algo}
The detailed implementation of CollabCoder is provided as the
pseudo-code in Algorithm~\ref{alg:collabcoder_coevolve}.


\begin{algorithm}[H]
\caption{CollabCoder pipeline}
\label{alg:collabcoder_coevolve}
\small
\begin{algorithmic}[1]
\Require Problem specification $P$, coding template $\mathcal{T}$, maximum number of iterations $T$, 
test set $\mathcal{D}=\{(x_i,y_i)\}_{i=1}^Q$, trust weights $\mathcal{W}=\{w_\pi, w_c, w_{\text{align}}\}$, 
execution oracle $\mathcal{O}$

\Ensure Final program $c$

\State $R^{(0)} \gets \emptyset$ \Comment{Reasoning Trajectory}
\State $\pi^{(0)} \gets \mathcal{A}_{\text{plan}}(P)$
\State $c^{(0)} \gets \mathcal{A}_{\text{code}}(\pi^{(0)}, \mathcal{T})$

\For{$t = 0$ \textbf{to} $T-1$}
    \State $\mathcal{F}^{(t)} \gets \mathcal{O}(c^{(t)})$ \Comment{Execute code}
    \If{\textsc{Satisfy}$(\mathcal{F}^{(t)})$} 
        \State \Return $c^{(t)}$
    \EndIf

    \State $(\mathcal{E}_\pi^{(t)}, \mathcal{E}_c^{(t)}, \mathcal{E}_{\text{align}}^{(t)}) \gets \mathcal{A}_{\text{CDM}}(P,\pi^{(t)},c^{(t)},\mathcal{F}^{(t)})$

\State $\Phi^{(t)}=\{\phi_{\{\pi,c,\text{align}\} \setminus \{i\},d}^{(t)} \mid i\in\{\pi,c,\text{align}\},\, d\in\{\text{plan},\text{code}\}\},\;
       \Psi^{(t)}=\{\varphi_{i,d}^{(t)} \mid i\in\{\pi,c,\text{align}\},\, d\in\{\text{plan},\text{code}\}\}
\gets \mathcal{A}_{\text{CDM}}(\mathcal{E}_\pi^{(t)},\mathcal{E}_c^{(t)},\mathcal{E}_{\text{align}}^{(t)})$

\State $D^{(t)} \gets 
\arg\max_{d \in \{\text{plan},\text{code}\}}
\sum_{i \in \{\pi,c,\text{align}\}}
w_i \cdot \phi_{\{\pi,c,\text{align}\}\setminus \{ i\},d}^{(t)} \cdot \varphi_{i,d}^{(t)}$

    \If{$D^{(t)} = \text{plan}$}
        \State $X^{(t)} \gets \pi^{(t)}$
        \State $\mathcal{E}_X^{(t)} \gets \mathcal{E}_\pi^{(t)}$
    \Else
        \State $X^{(t)} \gets c^{(t)}$
        \State $\mathcal{E}_X^{(t)} \gets \mathcal{E}_c^{(t)}$
    \EndIf

    \State $R^{(t+1)} \gets \mathcal{A}_{\text{RT}}(R^{(t)},\mathcal{E}_X^{(t)},P,X^{(t)},\mathcal{F}^{(t)})$

    \If{$D^{(t)} = \text{plan}$}
        \State $\pi^{(t+1)} \gets \mathcal{A}_{\text{plan}}(P,\pi^{(t)},\mathcal{F}^{(t)},R^{(t+1)})$
        \State $c^{(t+1)} \gets \mathcal{A}_{\text{code}}(\pi^{(t+1)},\mathcal{T})$
    \Else
        \State $\pi^{(t+1)} \gets \pi^{(t)}$
        \State $c^{(t+1)} \gets \mathcal{A}_{\text{code}}(P,c^{(t)},\mathcal{F}^{(t)},R^{(t+1)}, \mathcal{T})$
    \EndIf
\EndFor

\State \Return $c^{(T)}$
\end{algorithmic}
\end{algorithm}


\subsection{Prompt Templates}
For better reproducibility, we present all prompt
templates shown in Figures~\ref{fig:prompt-plan-init}–\ref{fig:prompt-code-refine} in the appendix.
\label{app:prompt}

\newtcolorbox{promptbox}{
  colback=green!10,
  colframe=black,
  boxrule=0.8pt,
  arc=1mm,
  left=4mm,
  right=4mm,
  top=3mm,
  bottom=3mm
}
\begin{figure*}[t]
\centering
\begin{promptbox}
\textbf{\Large INITIAL PLANNING ($\mathcal{A}_{\text{plan}}$)}


\textbf{Task:}\\
Generate a detailed step-by-step plan to solve the given programming problem. The plan should describe the reasoning and algorithmic approach without generating any executable code.


\begin{itemize}[itemsep=0.01pt, topsep=5pt]
  \item Recall a relevant but distinct example problem.
  \item Describe its solution approach and underlying algorithm.
  \item Based on this reasoning, produce a detailed plan for the original problem.
  \item Do \textbf{not} generate any code.
\end{itemize}


\textbf{Problem:}\\
\{\{problem\}\}


\textbf{Sample Test Cases:}\\
\{\{sample\_io\}\}

\end{promptbox}
\caption{Prompt for initial planning $\pi^{(0)} \gets \mathcal{A}_{\text{plan}}(P)$ (Algorithm~\ref{alg:collabcoder_coevolve}, Line~2).}
\label{fig:prompt-plan-init}
\end{figure*}

\begin{figure*}[t]
\centering
\begin{promptbox}
\textbf{\Large INITIAL CODE GENERATION ($\mathcal{A}_{\text{code}}$)}


\textbf{Task:}\\
Generate an executable program that solves the given problem by strictly following the provided plan. The code must conform to the specified programming language and input/output format.


\begin{itemize}[itemsep=0.01pt, topsep=5pt]
  \item You are given a step-by-step plan $\pi^{(0)}$ describing how to solve the problem.
  \item Implement the solution strictly according to this plan.
  \item If available, follow the provided coding template $\mathcal{T}$ without modification.
  \item Follow the sample input/output format exactly.
  \item Do \textbf{not} add extra explanations, comments outside the code, or auxiliary text.
  \item Do \textbf{not} include assertion or testing statements.
\end{itemize}


\textbf{IMPORTANT INSTRUCTIONS:}
\begin{itemize}[itemsep=0.01pt, topsep=5pt]
  \item The generated code must be written in \{\{language\}\}.
  \item The entire code must be enclosed within a triple backtick (\texttt{```}) block.
  \item Read input from standard input and write output to standard output.
  \item Do not include any extra print statements.
\end{itemize}


\textbf{Problem:}\\
\{\{problem\}\}


\textbf{Plan ($\pi^{(0)}$):}\\
\{\{plan\}\}


\textbf{Sample Test Cases:}\\
\{\{sample\_io\}\}

\end{promptbox}
\caption{Prompt for initial code generation 
$c^{(0)} \gets \mathcal{A}_{\text{code}}(\pi^{(0)}, \mathcal{T})$ (Algorithm~\ref{alg:collabcoder_coevolve}, Line~3).}
\label{fig:prompt-code-init}
\end{figure*}

\begin{figure*}[t]
\centering
\small
\begin{promptbox}
\textbf{\Large MERGED DIAGNOSTIC ANALYSIS ($\mathcal{A}_{\text{CDM}}$)}


\textbf{Task:}\\
Perform a comprehensive diagnostic analysis of the current plan, code, and their alignment with the problem, based on observed execution failures. The goal is to identify errors, inconsistencies, and misalignments without proposing new solutions.


\textbf{Context Provided:}
\begin{itemize}[itemsep=0.01pt, topsep=5pt]
  \item The original problem description $P$.
  \item The current plan $\pi^{(t)}$.
  \item The current code implementation $c^{(t)}$.
  \item The failure log $\mathcal{F}^{(t)}$, which records incorrect behavior on test cases.
\end{itemize}


\textbf{Response Structure (Strict):}
\begin{itemize}[itemsep=0.01pt, topsep=5pt]
  \item \textbf{Plan Analysis}
  \begin{itemize}
    \item \textbf{Simulation:} Step-by-step simulation of the plan on the failing test cases.
    \item \textbf{Insight:} Determine whether the plan is incorrect, or whether errors arise from plan-to-code translation, and explain how the plan should be corrected.
  \end{itemize}

  \item \textbf{Code Analysis}
  \begin{itemize}
    \item \textbf{Simulation:} Line-by-line execution of the code on the failing test cases.
    \item \textbf{Insight:} Identify implementation bugs or logical errors and explain how they should be fixed.
  \end{itemize}

  \item \textbf{Content Analysis}
  \begin{itemize}
    \item Provide a \emph{single concise insight} (4--5 sentences) evaluating the alignment between the problem, plan, and code.
    \item Conclude which component(s) should be updated (plan, code, both).
  \end{itemize}
\end{itemize}


\textbf{IMPORTANT:}
\begin{itemize}[itemsep=0.01pt, topsep=5pt]
  \item The failure log is always correct and must not be questioned.
  \item Do \textbf{not} generate new plans or code.
  \item Do \textbf{not} introduce alternative solutions.
  \item Strictly follow the specified structure.
\end{itemize}


\textbf{Problem ($P$):}\\
\{\{problem\}\}


\textbf{Current Plan ($\pi^{(t)}$):}\\
\{\{plan\}\}


\textbf{Current Code ($c^{(t)}$):}\\
\{\{code\}\}


\textbf{Failure Log ($\mathcal{F}^{(t)}$):}\\
\{\{failure\_log\}\}

\end{promptbox}
\caption{Merged diagnostic analysis 
$(\mathcal{E}_\pi^{(t)}, \mathcal{E}_c^{(t)}, \mathcal{E}_{\text{align}}^{(t)}) \gets 
\mathcal{A}_{\text{CDM}}(P, \pi^{(t)}, c^{(t)}, \mathcal{F}^{(t)})$
(Algorithm~\ref{alg:collabcoder_coevolve}, line~9).}
\label{fig:prompt-diag}
\end{figure*}

\begin{figure*}[t]
\centering
\small
\begin{promptbox}
\textbf{\Large CDM SCORING ($\mathcal{A}_{\text{CDM}}$)}
\small


\textbf{Task:}\\
Evaluate each candidate decision using both \emph{confidence} and \emph{consistency} criteria, based on diagnostic insights produced by multiple analysis agents. The goal is to quantitatively assess which decision should be taken in the next iteration.


\textbf{Context Provided:}
\begin{itemize}[itemsep=0.01pt, topsep=5pt]
  \item A set of candidate decisions (e.g., \emph{update plan}, \emph{update code only}).
  \item Diagnostic insights from multiple analysis types:
  \begin{itemize}
    \item Plan analysis
    \item Code analysis
    \item Content (alignment) analysis
  \end{itemize}
  \item Predefined analysis pairs for consistency evaluation.
\end{itemize}


\textbf{Scoring Definitions:}
\begin{itemize}[itemsep=0.01pt, topsep=5pt]
  \item \textbf{Confidence ($\Phi^{(t)}$):} Measures how strongly a single analysis supports or refutes a given decision.
  \item \textbf{Consistency ($\Psi^{(t)}$):} Measures the degree of agreement between pairs of analyses regarding the same decision.
\end{itemize}


\textbf{Scoring Rules:}
\begin{itemize}[itemsep=0.01pt, topsep=5pt]
  \item All scores must lie in the range $[0, 1]$.
  \item Higher confidence indicates stronger, more direct evidence.
  \item Higher consistency indicates stronger agreement across analyses.
  \item Contradictory insights must result in low scores.
\end{itemize}


\textbf{Response Format (Strict JSON):}
\begin{verbatim}
{
  "confidence_scores": {
    "<decision>": {
      "<analysis_type>": {
        "confidence": float,
        "reasoning": string
      }
    }
  },
  "consistency_scores": {
    "<decision>": {
      "<analysis1>-<analysis2>": {
        "consistency": float,
        "reasoning": string
      }
    }
  }
}
\end{verbatim}


\textbf{IMPORTANT:}
\begin{itemize}
  \item Output \textbf{JSON only}; do not include markdown or extra text.
  \item Use concise reasoning (1--3 sentences per score).
  \item If analyses contradict each other, assign low scores.
\end{itemize}


\textbf{Decisions:}\\
\{\{decisions\}\}


\textbf{Diagnostic Insights:}\\
\{\{analyses\}\}

\end{promptbox}
\caption{CDM scoring of confidence and consistency
$(\Phi^{(t)}, \Psi^{(t)}) \gets \mathcal{A}_{\text{CDM}}(\mathcal{E}_\pi^{(t)},\mathcal{E}_c^{(t)},\mathcal{E}_{\text{align}}^{(t)})$
(Algorithm~\ref{alg:collabcoder_coevolve}, line~10).}
\label{fig:prompt-score}
\end{figure*}

\begin{figure*}[t]
\centering
\begin{promptbox}
\small
\textbf{\Large REASONING TRAJECTORY UPDATE ($\mathcal{A}_{\text{RT}}$)}


\textbf{Task:}\\
Update the persistent debugging strategy based on newly observed diagnostic evidence, while maintaining continuity with the previous strategy.


\textbf{Inputs:}
\begin{itemize}[itemsep=0.01pt, topsep=5pt]
  \item Previous strategy $R^{(t)}$
  \item Diagnostic evidence $\mathcal{E}_X^{(t)}$ for the selected target $X \in \{\pi, c\}$
  \item Problem description $P$
  \item Current target state $X^{(t)}$
  \item Failure log $\mathcal{F}^{(t)}$
\end{itemize}


\textbf{Guidelines:}
\begin{itemize}[itemsep=0.01pt, topsep=5pt]
  \item Incorporate new evidence without repeating $R^{(t)}$ verbatim.
  \item State concrete next hypotheses or actions.
  \item Avoid ineffective or repeated fixes.
  \item Do \textbf{not} generate code or a new plan.
\end{itemize}


\textbf{Output:}\\
Return only the updated debugging strategy text $R^{(t+1)}$.

\end{promptbox}
\caption{Reasoning trajectory update
$R^{(t+1)} \gets \mathcal{A}_{\text{RT}}(R^{(t)}, \mathcal{E}_X^{(t)}, P, X^{(t)}, \mathcal{F}^{(t)})$
(Algorithm~\ref{alg:collabcoder_coevolve}, line~19).}
\label{fig:prompt-rt}
\end{figure*}

\begin{figure*}[t]
\centering
\begin{promptbox}
\small
\textbf{\Large PLAN REFINEMENT ($\mathcal{A}_{\text{plan}}$)}


\textbf{Task:}\\
Refine the current plan based on observed failures and the updated reasoning trajectory, producing a corrected plan for the next iteration.


\textbf{Inputs:}
\begin{itemize}[itemsep=0.01pt, topsep=5pt]
  \item Problem description $P$
  \item Current plan $\pi^{(t)}$
  \item Failure log $\mathcal{F}^{(t)}$
  \item Updated debugging strategy $R^{(t+1)}$
\end{itemize}


\textbf{Guidelines:}
\begin{itemize}[itemsep=0.01pt, topsep=5pt]
  \item Modify the plan to address diagnosed errors.
  \item Ensure logical coherence and step-by-step correctness.
  \item Do \textbf{not} generate executable code.
  \item Output the plan only, without explanations.
\end{itemize}


\textbf{Output:}\\
Return the updated plan $\pi^{(t+1)}$.

\end{promptbox}
\caption{Plan refinement
$\pi^{(t+1)} \gets \mathcal{A}_{\text{plan}}(P, \pi^{(t)}, \mathcal{F}^{(t)}, R^{(t+1)})$
(Algorithm~\ref{alg:collabcoder_coevolve}, line~21).}
\label{fig:prompt-plan-refine}
\end{figure*}

\begin{figure*}[t]
\centering
\begin{promptbox}
\small
\textbf{\Large CODE GENERATION AFTER PLAN UPDATE ($\mathcal{A}_{\text{code}}$)}


\textbf{Task:}\\
Generate a new code implementation based on the refined plan, incorporating guidance from the updated reasoning trajectory.


\textbf{Inputs:}
\begin{itemize}[itemsep=0.01pt, topsep=5pt]
  \item Refined plan $\pi^{(t+1)}$
  \item Coding template $\mathcal{T}$
\end{itemize}


\textbf{Guidelines:}
\begin{itemize}[itemsep=0.01pt, topsep=5pt]
  \item Implement the solution strictly following $\pi^{(t+1)}$.
  \item Respect the coding template defined by $\mathcal{T}$.
  \item Generate a \emph{new} implementation (do not reuse previous code).
  \item Do \textbf{not} include explanations or extra text.
\end{itemize}


\textbf{Output:}\\
Return the generated code $c^{(t+1)}$ only.

\end{promptbox}
\caption{Code generation after plan update
$c^{(t+1)} \gets \mathcal{A}_{\text{code}}(\pi^{(t+1)}, \mathcal{T})$
(Algorithm~\ref{alg:collabcoder_coevolve}, line~22).}
\label{fig:prompt-code-after-plan}
\end{figure*}
\begin{figure*}[t]
\centering
\begin{promptbox}
\small
\textbf{\Large CODE REFINEMENT / PATCHING ($\mathcal{A}_{\text{code}}$)}


\textbf{Task:}\\
Refine the existing code implementation to correct observed failures, guided by diagnostic insights and the updated reasoning trajectory.


\textbf{Inputs:}
\begin{itemize}[itemsep=0.01pt, topsep=5pt]
  \item Problem description $P$
  \item Current code $c^{(t)}$
  \item Coding template $\mathcal{T}$
  \item Failure log $\mathcal{F}^{(t)}$
  \item Updated debugging strategy $R^{(t+1)}$
\end{itemize}


\textbf{Guidelines:}
\begin{itemize}[itemsep=0.01pt, topsep=5pt]
  \item Modify the code to address diagnosed errors.
  \item Incorporate guidance from $R^{(t+1)}$.
  \item Respect the coding template defined by $\mathcal{T}$.
  \item Do \textbf{not} reuse the same incorrect implementation.
  \item Do \textbf{not} add testing or assertion code.
  \item Output only the corrected code.
\end{itemize}


\textbf{Output:}\\
Return the refined code $c^{(t+1)}$ enclosed in a code block.

\end{promptbox}
\caption{Code refinement / patching
$c^{(t+1)} \gets \mathcal{A}_{\text{code}}(P,  c^{(t)}, \mathcal{F}^{(t)}, R^{(t+1)},\mathcal{T})$
(Algorithm~\ref{alg:collabcoder_coevolve}, line~25).}
\label{fig:prompt-code-refine}
\end{figure*}

\end{document}